\documentclass{article}

\usepackage{color}
\usepackage{graphicx, graphics}
\usepackage{subfigure}
\usepackage{amsmath,amstext}  
\usepackage{amsfonts,amssymb, bbm,natbib} 
\usepackage[LGR, T1]{fontenc}
\usepackage[utf8]{inputenc}
\usepackage[english]{babel}
\usepackage{epsfig}
\usepackage{epsf}
\usepackage{vmargin}

\definecolor{dkg}{rgb}{0,0.5,0}

\newcommand{\ind}[1]{\mathbbm{1}_{\{#1\}}}
\newcommand{\indnex}{\mathbbm{1}_{\overline{\cE}}}
\newcommand{\indnexj}{\mathbbm{1}_{\overline{\cE}_j}}
\newcommand{\bs}[1]{\mathbf{#1}}
\newcommand{\dE}{\mathbb{E}}
\newcommand{\dF}{\mathbb{F}}
\newcommand{\dG}{\mathbb{G}}
\newcommand{\dP}{\mathbb{P}}
\newcommand{\dR}{\mathbb{R}}

\newcommand{\dT}{\mathbb{T}}
\newcommand{\wh}{\widehat}
\newcommand{\wt}{\widetilde}
\newcommand{\rI}{\bs{{I}}}

\newcommand{\veps}{\varepsilon}
\newcommand{\cA}{\mathcal{A}}

\newcommand{\cE}{\mathcal{E}}
\newcommand{\cF}{\mathcal{F}}
\newcommand{\cG}{\mathcal{G}}

\newcommand{\cL}{\mathcal{L}}
\newcommand{\cN}{\mathcal{N}}
\newcommand{\cO}{\mathcal{O}}

\newcommand{\reff}[1]{(\ref{#1})}

\title{Statistical study of asymmetry in cell lineage data} 
\author{ Beno\^{\i}te de Saporta\\
Univ. Bordeaux, Gretha, UMR 5113, IMB, UMR 5251, F-33400 Talence, France\\
CNRS, Gretha, UMR 5113, IMB, UMR 5251, F-33400 Talence, France\\
INRIA Bordeaux Sud Ouest, team CQFD, F-33400 Talence, France\\
\and Anne G\'egout-Petit\\
Univ. Bordeaux, IMB, UMR 5251, F-33400 Talence, France\\
CNRS, IMB, UMR 5251, F-33400 Talence, France\\
INRIA Bordeaux Sud Ouest, team CQFD, F-33400 Talence, France\\
\and Laurence Marsalle \\
Univ. Lille 1, Laboratoire Paul Painlev\'e, UMR~8524, F-59 655 Villeneuve d'Ascq, France\\
CNRS, Laboratoire Paul Painlev\'e, UMR~8524, F-59 655 Villeneuve d'Ascq, France}

\begin{document}

\newtheorem{lemma}{Lemma}[section]
\newtheorem{proposition}[lemma]{Proposition}
\newtheorem{theorem}[lemma]{Theorem}
\newtheorem{cor}[lemma]{Corollary}

\maketitle

\begin{abstract}
{A} rigorous methodology {is proposed} to study cell division data consisting in several observed genealogical trees of possibly different shapes. {The} procedure take{s} into account missing observations, data from different trees, as well as the dependence structure within genealogical trees. {Its main {new} feature is the joint use of} all available information {from several data sets instead of single data set estimation, to avoid} the drawbacks of low accuracy for estimators or low power for tests on small single-trees.
The data {is modeled} by an asymmetric bifurcating autoregressive process and possibly missing observations {are taken into account} by modeling the genealogies with a two-type Galton-Watson process. Least-squares estimators of the unknown parameters of the processes {are given} and symmetry tests {are derived}. Results are applied on real data of Escherichia coli division {and an empirical study of the convergence rates of the estimators and power of the tests is conducted on simulated data}.
\end{abstract}

\section{Introduction}
\label{s:intro}
Cell lineage data consist of observations of some quantitative characteristic of the cells (e.g. their length, growth rate, {time until division}, \ldots) over several generations descended from an initial cell. Track is kept of the genealogy to study the inherited effects on the evolution of the characteristic. As a cell usually gives birth to two offspring by division, such genealogies are structured as binary trees. \cite{CoSt86} first adapted autoregressive processes to this binary tree structure by introducing bifurcating autoregressive processes (BAR). This parametric model takes into account both the environmental and inherited effects. Inference on this model has been proposed based on either a single-tree growing to infinity, see e.g. \cite{CoSt86}, \cite{Hug96}, \cite{HuBa00}, \cite{ZhBa05b} or for an asymptotically infinite number of small replicated trees, see e.g. \cite{HuSt94}, \cite{HuBa99}.

More recently, studies of aging in single cell organisms by \cite{SMT05} suggested that cell division may not be symmetric. An asymmetric BAR model was therefore proposed by \cite{Guy07}, where the two sets of parameters corresponding to sister cells are allowed to be different. Inference for this model was only investigated for single-trees growing to infinity, see \cite{Guy07}, \cite{BSG09} for the fully observed model or \cite{DM08}, \cite{SGM11}, \cite{SPL12} {for} missing data {models}. 

Cell division data often consist in recordings over several genealogies of cells evolving in similar experimental conditions. For instance, \cite{SMT05} filmed 94 colonies of Escherichia coli cells dividing between four and nine times. 
We therefore propose a new rigorous approach to take into account all the available information. Indeed, we propose an inference based on a finite fixed number of replicated trees when the total number of observed cells tends to infinity. We use the missing data asymmetric BAR model introduced by \cite{SGM11}. In this approach, the observed genealogies are modeled with a two-type Galton-Watson (GW) process. However, we propose a different least-squares estimator for the parameters of the BAR process that does not correspond to the single-tree estimators averaged on the replicated trees. We also propose an estimator of the parameters of the GW process specific to our binary tree structure and not based simply on the observation of the number of cells of each type in each generation as in \cite{Gut91}, \cite{MaaTou05}. 
We study the consistency and asymptotic normality of our estimators and derive asymptotic confidence intervals as well as Wald's type tests to investigate the asymmetry of the data for both the BAR and GW processes. Our results are applied to the Escherichia coli data of \cite{SMT05}. {We also provide an empirical study of the convergence rate of our estimators and of the power of the symmetry tests on simulated data.}

The paper is organized as follows. {In Section \ref{s:methodo}, we describe a methodology for} {least-squares estimation based on multiple data sets} {in a general framework}. In Section~\ref{s:model}, we present the BAR and observations models. In Section~\ref{s:inference} we give our estimators and state their asymptotic properties. In Section~\ref{s:data}, we propose a new investigation of \cite{SMT05} data. In Section~\ref{s:simu} we give simulation results. The precise statement of the convergence results, the explicit form of the asymptotic variance of the estimators and the convergence proofs are postponed to the appendix.

%
\section{Methodology}
\label{s:methodo}
We work with the following general framework. Consider that several data sets are available{,} obtained in similar experimental conditions and {then} {assumed to} come from the same parametric model. Suppose that {there exists} a consistent least-squares estimator for the parametric model. This estimator can be computed on each individual data set, but we would like to take into account all the data at disposal, which should improve the accuracy of the estimation.

To this aim, we assume that the different data sets are independent realizations of the parametric model. A natural idea is to average the single-set estimators. It may be a good approach if the single-set estimators have roughly the same variance, which is usually the case when the data sets have the same size. However, if the data sets have very different sizes, the single-set estimators may have variances of different orders and this {direct} approach becomes dubious.

Instead, we propose to use a global least-squares estimator. Suppose that we have $m$ data sets. Let $\bs{\theta}$ be the ({possibly} multivariate) parameter to be estimated, and $\wh{\bs{\theta}}_{j,n}$ the least-squares estimator build with the $j$-th data set for $1\leq j\leq m$. The global least-squares estimator $\wh{\bs{\theta}}_{n}$ decomposes as
\begin{equation*}
\wh{\bs{\theta}}_{n}=\Big(\sum_{j=1}^m\bs{\Sigma}_{j,n}\Big)^{-1}\sum_{j=1}^m\bs{V}_{j,n},
\end{equation*}
where $\bs{\Sigma}_{j,n}$ is a normalizing matrix and $\bs{V}_{j,n}$ a vector of the same size as $\bs{\theta}$,  involved in the decomposition of the single-set least-squares estimator $\wh{\bs{\theta}}_{j,n}$ as follows
\begin{equation*}
\wh{\bs{\theta}}_{j,n}=\bs{\Sigma}_{j,n}^{-1}\bs{V}_{j,n}.
\end{equation*}
Note that {the estimator} $\wh{\bs{\theta}}_{n}$ thus constructed is neither an average nor a function of the $\wh{\bs{\theta}}_{j,n}$. Hence, the asymptotic behavior of the global estimator $\wh{\bs{\theta}}_{n}$ cannot be deduced from that of the single-set estimators $\wh{\bs{\theta}}_{j,n}$. Nevertheless, the asymptotic behavior of $\wh{\bs{\theta}}_{j,n}$ is often obtained through the convergence of the normalizing matrices $\bs{\Sigma}_{j,n}$ and of the vectors $\bs{V}_{j,n}$ {separately}, {which gives} the convergence of the global estimator $\wh{\bs{\theta}}_{n}$ as the number $m$ of data sets is fixed. Note that the asymptotic is not the number  $m$ of data sets.

{The aim of this paper is to} apply this methodology to cell division data {with missing data}. In this special case, the convergence of the global estimator $\wh{\bs{\theta}}_{n}$ is not straightforward, because we have to prove it on a set where the convergence of each $\bs{\Sigma}_{j,n}$ and $\bs{V}_{j,n}$ is not ensured.

\color{black}
%
\section{Model}
\label{s:model}
Our aim is to estimate the parameters of coupled BAR and GW processes through $m$ i.i.d. realizations of the processes. We first define our parametric model and introduce our notation{s}. The BAR and GW processes have the same dynamics as in \cite{SGM11}, the main difference is that our inference is here based on several i.i.d. realizations of the processes, instead of a single one. Additional notation{s} together with the precise technical assumptions are specified in \ref{s:apx1}. 
\subsection{Bifurcating autoregressive model}
\label{ss:BAR}
Consider $m$ i.i.d. replications of the asymmetric BAR process with coefficient {$\bs{\theta}=$}${(a_0, b_0, a_1, b_1)}\in\dR^4$. More precisely, for $1\leq j\leq m$, the first cell in genealogy $j$ is labelled $(j,1)$ and for $k\geq 1$, the two offspring of cell $(j,k)$ are labelled $(j,2k)$ and $(j,2k+1)$. As we consider an asymmetric model, each cell has a \emph{type} defined by its label: $(j,2k)$ has type \emph{even} and $(j,2k+1)$ has type \emph{odd}. The characteristic of cell $k$ in genealogy $j$ is denoted by $X_{(j,k)}$.  The BAR processes are defined recursively as follows: for all $ 1 \leq j \leq m$ and $k\geq 1$, one has
\begin{equation}\label{defbar}
\left\{
    \begin{array}{lcccccl}
     X_{(j, 2k)} & = & {{a_0}} &+ &{{b_0}}X_{(j, k)} &+ &\varepsilon_{(j,2k)}, \\
     X_{(j,2k+1)} & = & {{a_1}} & +& {{b_1}}X_{(j, k)} &+ &\varepsilon_{(j,2k+1)}.
    \end{array}\right.
\end{equation}
Let {us also define the variance and covariance of the noise sequence}
$${\sigma^2_0=\dE[\varepsilon_{(j,2k)}^2],\qquad \sigma_1^2=\dE[\varepsilon_{(j,2k+1)}^2],}\qquad
\rho=\dE[\varepsilon_{(j,2k)}\varepsilon_{(j,2k+1)}].$$ 
Our goal is to estimate the parameters {$\bs{\theta}=$}${(a_0,b_0,a_1,b_1)}$ and 
$({\sigma_0^2,\sigma^2_1}
, \rho)$, and then test if ${(a_0,b_0)=(a_1,b_1)}$ or not.  
%
\subsection{Observation process}
\label{ss:GW}
\begin{figure}
 \centerline{\includegraphics[height=5cm]{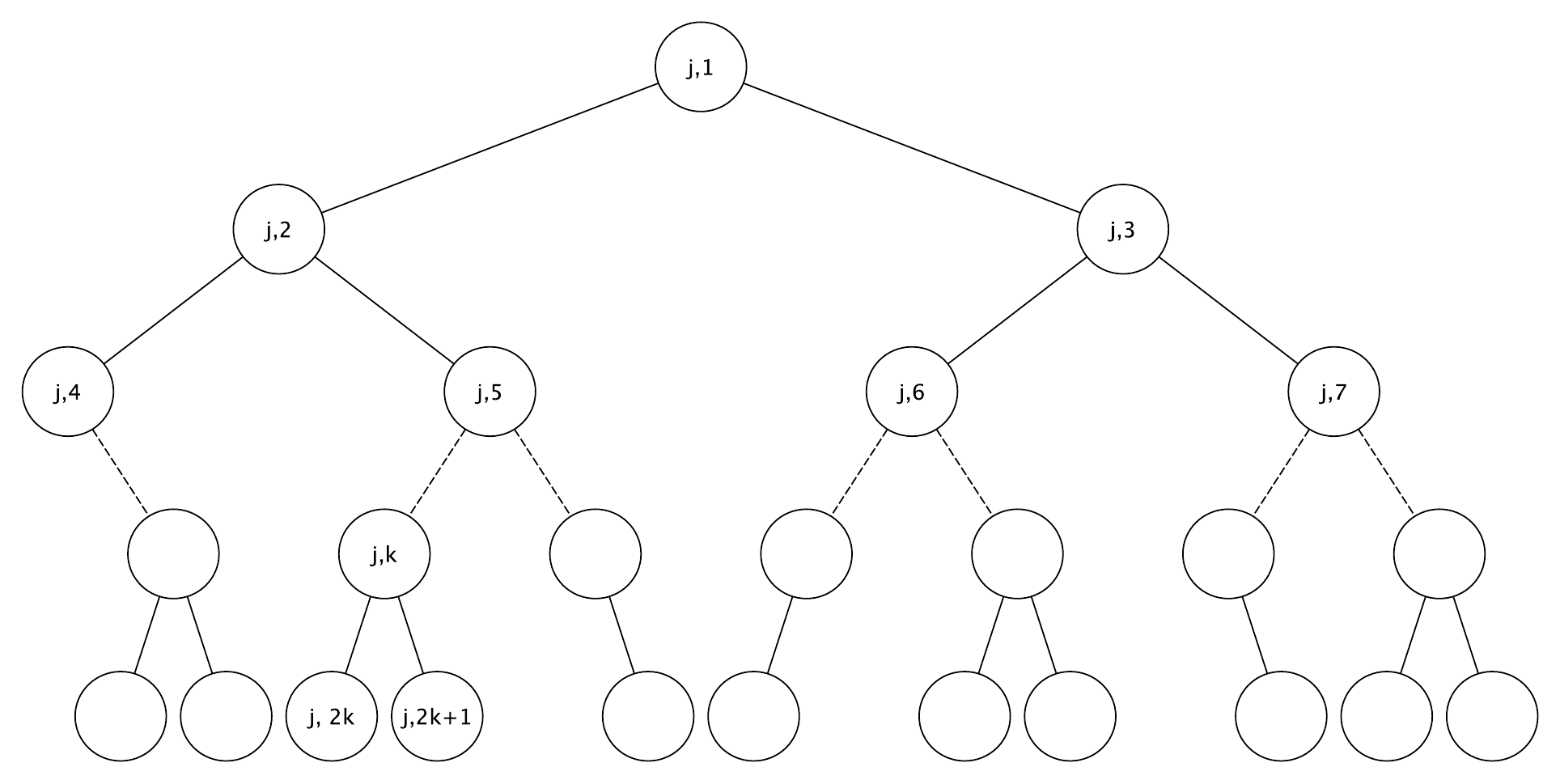}}
\caption{A tree of observed cells.}
\label{f:tree}
\end{figure}

We now turn to the observation process $(\delta_{(j,k)})$ that encodes for the presence or absence of cell measurements in the available data
\begin{equation*}
\delta_{(j,k)}=\begin{cases}
1 & \text{if cell $k$ in genealogy $j$ is observed,}\\
0 & \text{if cell $k$ in genealogy $j$ is not observed.}
\end{cases}
\end{equation*}
To take into account possible asymmetry in the observation process, we use a two-type Galton-Watson model.  The relevance of this model to E. coli data is discussed in section~\ref{s:data}.
Again, we suppose all the $m$ observation processes to be drawn independently from the same two-type GW process. More precisely, for all $1\leq j\leq m$, we {model} the observation process $(\delta_{(j,k)})_{k\geq1}$ for the $j$-th genealogy as follows. We set $\delta_{(j,1)}=1$ and {draw} $(\delta_{(j,2k)},\delta_{(j,2k+1)})$ independently from one another with a law depending on the type of cell $k$. More precisely, for $i \in \{0,1\}$, if $k$ is of type $i$ {we set}
$$\dP\Big((\delta_{(j,2k)},\delta_{(j,2k+1)})=(l_0,l_1) \ \Big|\  \delta_{(j,k)}=1\Big) = p^{(i)}(l_0,l_1),$$
$${\dP\Big((\delta_{(j,2k)},\delta_{(j,2k+1)})=(0,0) \ \Big|\  \delta_{(j,k)}=0\Big) = 1,}$$
for all $(l_0,l_1) \in \{0,1\}^2$. Thus, $p^{(i)}(l_0,l_1)$ is the probability that a cell of type $i$ has $l_0$ daughter of type $0$ and $l_1$ daughter of type $1$. {And if a cell is missing, its descendants are missing as well}. Figure \ref{f:tree} gives an example of realization of an observation process.
We also assume that the observation processes are independent from the BAR processes.
%
\section{Inference}
\label{s:inference}
Our first goal is to estimate the reproduction probabilities $p^{(i)}(l_0,l_1)$ of the GW process from the $m$ genealogies of observed cells up to the $n$-th generation to be able to test the symmetry of the GW model itself. Our second goal is to estimate ${\bs{\theta}}=(a_0,b_0,a_1,b_1)^t$ from all the observed individuals of the $m$ trees up to the $n$-th generation. We then give the asymptotic properties of our estimator to be able to build confidence intervals and symmetry tests for ${\bs{\theta}}$.

Denote by $|\dT_n^*|$ the total number of observed cells in the $m$ trees up to the $n$-th generation of offspring from the original ancestors, and let
\begin{equation*}
\overline{\cE}=\{\lim_{n\rightarrow\infty}|\dT_n^*| = \infty\}
\end{equation*}
be the non-extinction set, on which the global cell population grows to infinity. {Thus, our asymptotic results only hold on the set $\overline{\cE}$. This global non-extinction set is the union and not the intersection of the non-extinction sets of each single-tree. It means that some trees may extinct, which allows us to take into account trees with a different number of observed generations.} {We are thus in a case where averaging single-tree estimators is not recommended. The possibility of extinction for some trees is also the reason why the convergence of the multiple-trees estimator $\wh{\bs{\theta}}_{n}$ is not straightforward from existing results in the literature.}
Conditions for the probability of non-extinction to be positive are given in~\ref{s:apx1}.
\subsection{Estimation of the reproduction law of the GW process}
\label{ss:ES}
There are many references on the inference of a multi-type GW process, see for instance \citep{Gut91} and \citep{MaaTou05}. Our context of estimation is very specific because the information given by $(\delta_{(j,k)})$ is more precise than that given by the number of cells of each type in a given generation that is usually used in the literature. Indeed, not only do we know the number of cells of each type in each generation, but we also know their precise positions on the binary tree of cell division. The empiric estimators of the reproduction probabilities using data up to the $n$-th generation are then, for $i,l_0,l_1$ in $\{0,1\}$
\begin{eqnarray*}
{\widehat{p}_n^{(i)}(l_0,l_1) }
& = & \frac{\sum_{j=1}^m\sum_{k\in\dT_{n-2}}\delta_{(j,2k+i)}\phi_{l_0}(\delta_{(j,2(2k+i))})\phi_{l_1}(\delta_{(j,2(2k+i)+1)})}{\sum_{j=1}^m\sum_{k\in\dT_{n-2}}\delta_{(j,2k+i)}},
\end{eqnarray*}
where $\phi_0(x)=1-x$, $\phi_1(x)=x$, and if the denominator is non zero, the estimator equalling zero otherwise. Note that the numerator is just the number of cells of type $i$ in all the trees up to generation $n-1$ that have exactly $l_0$ daughter of type $0$ and $l_1$ daughter of type $1$ in the $n$-th generation. The denominator is the total number of cells of type $i$ in all the trees up to generation $n-1$.
Set also 
$$\bs{p}^{(i)}={(p^{(i)}(1,1),p^{(i)}(1,0),p^{(i)}(0,1),p^{(i)}(0,0))}^t,$$ 
the vector of the $4$ reproduction probabilities for a mother of type $i$, $\bs{p}=((\bs{p}^{(0)})^t, (\bs{p}^{(1)})^t)^t$ the vector of all $8$ reproduction probabilities and $\widehat{\bs{p}}_{n}$ its empirical estimator.
%
\subsection{Least-squares estimation for the BAR parameters}
\label{ss:LS}
For the parameters of the BAR process, we use the standard least-squares (LS) estimator $\wh{{\bs{\theta}}}_n$ {with all the available data from the $m$ trees up to generation $n$. It}
 minimizes 
$$\Delta_n(\bs{\theta})=\sum_{j=1}^m\sum_{k\in \dT_{n-1}}\delta_{(j,2k)}(X_{(j,2k)}-{a_0}-{b_0}X_{(j,k)})^2 + \delta_{(j,2k+1)} (X_{(j,2k+1)}-{a_1}-{b_1}X_{(j,k)})^2.$$ Consequently,  for all $n\geq 1$ we have $\wh{\bs{\theta}}_n={(\wh{a}_{0,n},\wh{b}_{0,n},\wh{a}_{1,n},\wh{b}_{1,n}})^t$ with
\begin{equation}\label{defLS}
\wh{\bs{\theta}}_n=  \bs{\Sigma}_{n-1}^{-1}\sum_{j=1}^m\sum_{k \in \mathbb{T}_{n-1}}
\left(
\delta_{(j,2k)}X_{(j,2k)},\
\delta_{(j,2k)}X_{(j,k)}X_{(j,2k)},\
\delta_{(j,2k+1)}X_{(j,2k+1)},\
\delta_{(j,2k+1)}X_{(j,k)}X_{(j,2k+1)}
\right)^t
\end{equation}
where, for $i\in\{0,1\}$ {we defined}
\begin{equation*}
\bs{\Sigma}_{n} = \left( \begin{array}{cc}
\bs{S}^0_{n} & 0 \\
0 & \bs{S}^1_{n}
\end{array} \right),\quad
\bs{S}^i_{n} =\sum_{j=1}^m \sum_{k \in \mathbb{T}_{n}}\delta_{(j,2k+i)}\left(
\begin{array}{cc}
1 & X_{(j,k)} \\
X_{(j,k)} &X^2_{(j,k)}
\end{array}\right).
\end{equation*}
Note that in the normalizing matrices $\bs{S}^i_{n}$ the sum is over all observed cells for which a daughter of type $i$ is observed, and not merely over all observed cells.
To estimate the variance parameters ${\sigma^2_i}$ and $\rho$, we define the empiric residuals.
For all $2^\ell\leq k\leq2^{\ell+1}-1$ and $1\leq j\leq m$ set
\begin{equation}
\label{defepschap}
\left\{
    \begin{array}{lccclcl}
     \wh{\veps}_{(j,2k)} &=& \delta_{(j,2k)}(X_{(j,2k)} &-& {\wh{a}_{0,\ell}} &-& {\wh{b}_{0,\ell}}X_{(j,k)}), \vspace{1ex}\\
     \wh{\veps}_{(j,2k+1)} &=& \delta_{(j,2k+1)}(X_{(j,2k+1)} &-& {\wh{a}_{1,\ell}} &-& {\wh{b}_{1,\ell}}X_{(j,k)}).
    \end{array}\right.
\end{equation}
We propose the following empirical estimators 
$${\wh{\sigma}^2_{i,n} = \frac{1}{ |\dT_{n-1}^{*i}|} \sum_{j=1}^m \sum_{k \in\dT_{n-1}} \wh{\veps}_{(j,2k+i)}^2},
\qquad 
\wh{\rho}_n = \frac{1}{|\dT_{n-1}^{*01}|} \sum_{j=1}^m \sum_{k \in\dT_{n-1}} \wh{\veps}_{(j,2k)} \wh{\veps}_{(j,2k+1)},$$ 
where { $|\dT_{n}^{*i}|$ is the set of all cells which have at least one offspring of type $i$, for $i\in\{0,1\}$ and} $|\dT_{n}^{*01}|$ is the set of all the cells which have exactly two offspring, in the $m$ trees up to generation $n$.
%
\subsection{Consistency and normality}
\label{ss:main}
We now state the convergence results we obtain for the estimators above. The assumptions (H.1) to (H.6) are given in~\ref{ss:joint model}. These results hold on the non-extinction set $\overline{\cE}$.  
\begin{theorem}\label{th:CVp}
Under assumptions \emph{ (H.5-6)} and for all $i$, $l_0$ and $l_1$ in $\{0,1\}$, $\widehat{p}_n^{(i)}(l_0,l_1)$ converges to $p^{(i)}(l_0,l_1)$ almost surely on $\overline{\cE}$.  
Under assumptions \emph{(H.0-6)},
$\wh{\bs{\theta}}_{n}$,  {$\wh{{\sigma}}_{0,n}^2$, $\wh{{\sigma}}_{1,n}^2$} and $\wh{{\rho}}_{n}$ converge to $\bs{\theta}$, {$\sigma^2_0$, $\sigma^2_1$} and $\rho$ respectively, almost surely on $\overline{\cE}$. 
\end{theorem}
The asymptotic normality results are only valid conditionally to the non-extinction of the global cell population.
\begin{theorem}\label{th:TCL}
Under assumptions \emph{ (H.5-6)} we have 
\begin{equation*}
\sqrt{|\dT_{n-1}^*|}(\widehat{\bs{p}}_{n}-\bs{p})
\xrightarrow{\cL}
\cN(0, \bs{V} ),
\end{equation*}
and under assumptions \emph{ (H.0-6)}, we have
\begin{eqnarray*}
\sqrt{|\dT^*_{n-1}|} (\widehat{\bs{\theta}}_{n}-\bs{\theta})\xrightarrow{\cL}\cN(0,\bs{\Gamma_{\theta}}),&& 
\sqrt{|\dT^*_{n}|} {(\wh{\sigma}^2_{0,n}-{\sigma}^2_0,\wh{\sigma}^2_{1,n}-{\sigma}^1_0)^t}\xrightarrow{\cL}\cN(0,{\bs{\Gamma_{\sigma}}}),\\
\sqrt{|\dT^{*01}_{n-1}|} (\wh{\rho}_n-{\rho})&\xrightarrow{\cL}&\cN(0,\gamma_{\rho}),
\end{eqnarray*}
conditionally to $\overline{\cE}$. The explicit form of the variance matrices $\bs{V}$, $\bs{\Gamma_{\theta}}$, {$\bs{\Gamma_{\sigma}}$} and of $\gamma_{\rho}$ is given in Eq.~(\ref{defV}), (\ref{defGamma}), (\ref{defgammasig}) and (\ref{defgammarho}) respectively.
\end{theorem}

The proofs of these results are given in~\ref{ss:apxGWconsist} and \ref{ss:apxGWnorm} for the GW process and in~\ref{ss:apxBARconsist} and \ref{ss:apxBARnorm} for the BAR process. From the asymptotic normality, one can naturally construct confidence intervals and tests. Their explicit formulas are given in~\ref{ss:CIGW} and \ref{ss:CIBAR}.
%
\section{Data analysis}
\label{s:data}
We applied our procedure to the Escherichia coli data of \cite{SMT05}. The biological issue addressed is aging in single cell organisms.  E. coli is a rod-shaped bacterium that reproduces by dividing in the middle. Each cell {has thus a new \emph{pole} (due to the division of its mother) and an old one (one of the two poles of its mother)}. The cell that inherits the old pole of its mother is called the \emph{old pole} cell, the other one is called the \emph{new pole} cell. Therefore, each cell has a \emph{type}: old pole or new pole cell , inducing asymmetry in the cell division. On a binary tree, the new pole cells are labelled by an even number and the old pole cells by an odd number. 

\cite{SMT05} filmed 94 colonies of dividing E. coli cells, determining the complete lineage and the growth rate of each cell. The number of divisions goes from four to nine. The 94 data sets gather {$|\dT_9^*|=22394$} data (11189 of type even and 11205 of type odd). Not a single data tree is complete. Missing data mainly do not come from cell death (only 16 cells are recorded to die) but from measurement difficulties due mostly to overlapping cells or cells wandering away from the field of view. Note also that for a growth rate to be recorded, the cell needs to be observed through its whole life cycle. If this is not the case, there is no record at all, so that a censored data model is not relevant. The observed average growth rate of even (resp. odd) cells is 0.0371 (resp.  0.0369). These data were investigated in \citep{SMT05, GBPSDT05, Guy07, SGM11, SPL12}. 

\cite{SMT05} proposed a statistical study of the averaged genealogy and pair-wise comparison of sister cells. They concluded that the old pole cells exhibit cumulatively slowed growth, less offspring biomass production and an increased probability of death whereas single-experiment analyses did not. However they assumed independence between the averaged couples of sister cells, which does not hold in such genealogies.

The other studies are based on single-tree analyses instead of averaging all the genealogical trees.
\citet{GBPSDT05} model the growth rate by a Markovian bifurcating process, but their procedure does not take into account the dependence between pairs of sister cells either. The asymmetry was rejected ({p-value$<0.1$}) in half of the experiments so that a global conclusion was difficult. \cite{Guy07} has then investigated the asymptotical properties of a more general asymmetric Markovian bifurcating autoregressive process, and he rigorously constructed a Wald's type test to study the asymmetry of the process.
However, his model does not take into account the possibly missing data from the genealogies. The author investigates the method on the 94 data sets but it is not clear how he manages missing data. More recently, \citet{SGM11} proposed a single-tree analysis with a rigorous method to deal with the missing data and carried out their analysis on the largest data set, concluding to asymmetry on this single set. Further single-tree studies of the 51 data sets issued from the 94 colonies containing at least 8 generations were conducted in \cite{SPL12}. The symmetry hypothesis is rejected in one set out of four for ${(a_0,b_0)=(a_1,b_1)}$ and one out of eight for ${a_0/(1-b_0)=a_1/(1-b_1)}$ forbidding a global conclusion. Simulation studies tend to prove that the power of the tests on single-trees is quite low for only eight or nine generations. This is what motivated the present study and urged us to use all the data available in one global estimation, rather than single-tree 
analyses.

In this section, we propose a new investigation of E. coli data of \citep{SMT05} where for the first time
 the dependence structure between cells within a genealogy is fully taken into account, 
 missing data are taken care of rigorously,
 all the available data, i.e. the 94 sets, are analyzed at once
 and both the growth rate and the number/type of descendants are investigated.
It is sensible to consider that all the data sets correspond to BAR processes with the same coefficients as the experiments where conducted in similar conditions. Moreover, a direct comparison of single-tree estimations would be meaningless as the data trees do not all have the same number of generations, and it would be impossible to determine whether variations in the computed single-tree estimators come from an intrinsic variability between trees or just the low accuracy of the estimators for small trees. The original estimation procedure described in {Section~\ref{s:methodo}} enables us to use all the information available without the drawbacks of low accuracy for estimators or low power for tests on small single-trees.
%
\subsection{Symmetry of the BAR process}
\label{ss:data1}
We now give the results of our new investigation of the E. coli growth rate data of \citep{SMT05}.
We suppose that the growth rate of cells in each lineage is modeled by the BAR process defined in Eq.~(\ref {defbar}) and observed through the two-type GW process defined in section \ref{ss:GW}. The experiments were independent and lead in the same conditions corresponding to independence and identical distribution of the processes  $(X_{(j,\cdot)}, \delta_{(j,\cdot)})$, $1 \leq j \leq m$.

We first give the point and interval estimation for the various parameters of the BAR process.
Table \ref{tab:1} gives the estimation $\wh{\bs{\theta}}_9$ of $\bs{\theta}$ with the 95\%  confidence interval (CI) of each coefficient {together with an estimation of $a_i/(1-b_i)$}. {This value is interesting in itself as }{$a_i/(1-b_i)$ is the fixed point of  the equation $\dE[X_{2k+i}]=a_i+b_i\dE[X_i]$. Thus it corresponds to the asymptotic mean growth rate of the cells in the lineage always inheriting the new pole from the mother ($i=0$) or always inheriting the old pole ($i=1$).}
The confidence intervals of ${b_0}$ and ${b_1}$ show that the \emph{non explosion} assumption $|{b_0}|<1$ and $|{b_1}|<1$ is satisfied. {Note that although the number of observed generations $n=9$ may seem too small to obtain the consistency of our estimators, Theorem \ref{th:TCL} shows that their variance is of order $|\dT_n^*|^{-1/2}$. Here the total number of observed cells is high enough as $|\dT_9^*|=22394$. In addition, an empirical study of the convergence rate on simulated data is conducted in the next section to validate that $9$ observed generations is enough.}
\begin{table}[htdp]  
\begin{tabular}{cccccc}
\hline\noalign{\smallskip}
parameter & estimation & CI&parameter & estimation & CI \\
\noalign{\smallskip}\hline\noalign{\smallskip}
${a_0}$&$0.0203$&${[0.0202  ; 0.0204]}$
&${a_1}$&$0.0195$&${[ 0.0194  ; 0.0196]}$\\
${b_0}$&$0.4615$&${[0.4417   ; 0.4812]}$
&${b_1}$&$0.4782$&${[0.4631 ; 0.4933]}$\\
${a_0/(1-b_0)}$&$0.03773$&${[0.03756   ; 0.03790]}$
&${a_1/(1-b_1)}$&$0.03734$&${[0.03717 ; 0.03752]}$\\
\noalign{\smallskip}\hline
\end{tabular}
\caption{Estimation and 95 \% CI of $\bs{\theta}$ {and $a_i/(1-b_i)$}.}
\label{tab:1}    
\end{table}

Table \ref{tab:2} gives the estimation{s $\wh{{\sigma}}^2_{i,9}$ of ${\sigma}^2_i$} and  $\wh{{\rho}}_9$ of ${\rho}$ with the 95\% CI of each coefficient. {The hypothesis of equality of variances $\sigma_0^2=\sigma_1^2$ is not rejected (p-value$=0.19$).} {From the biological point of view, t}{his result is not surprising as the noise sequence represents the influence of the environment and both sister cells are born and grow in the same local environment.}
\begin{table}[t]     
\begin{center}
\begin{tabular}{ccc}
\hline\noalign{\smallskip}
parameter & estimation & CI \\
\noalign{\smallskip}\hline\noalign{\smallskip}
${\sigma}^2_{{0}}$&${2.28\cdot10^{-5}}$&${[  0.88\cdot10^{-5}  ; 3.67\cdot10^{-5}]}$\\
${\sigma}^2_{{1}}$&${1.34\cdot10^{-5}}$&${[  1.29\cdot10^{-5}  ; 1.40\cdot10^{-5}]}$\\
${\rho}$&$0.48\cdot10^{-5}$&$[ 0.44\cdot10^{-5}  ; 0.52\cdot10^{-5}]$\\
\noalign{\smallskip}\hline
\end{tabular}
\caption{Estimation and 95 \% CI of ${\sigma}^2_{{i}}$ and ${\rho}$}
\label{tab:2} 
\end{center}
\end{table}
 
We now turn to the results of symmetry tests.
The hypothesis of equality of the couples ${(a_0,b_0)=(a_1,b_1)}$ is strongly rejected ({p-value $=10^{-5}$}).
The hypothesis of the equality of the two fixed points ${a_0/(1-b_0)}$ and ${a_1/(1-b_1)}$ of the BAR process is also rejected ({p-value $=  2\cdot10^{-3}$}). 
We can therefore rigorously confirm that there is a statistically significant asymmetry in the division of E. coli. 
{Biologically we can thus conclude that the growth rates of the old pole and new pole cells do have different dynamics. This is interpreted as aging for the single cell organism E. coli, see \cite{SMT05,Lydia10}.}
%
\subsection{Symmetry of the GW process}
\label{ss:data2}
Let us now turn to the asymmetry of the GW process itself. Note that to our best knowledge, it is the first time this question is investigated for the E. coli data of \citep{SMT05}. 
We estimated the parameters ${p}^{(i)}(l_0,l_1)$ of the reproduction laws of the underlying GW process. Table \ref{tab:3} gives the estimations $\wh{{p}}_9^{(i)}(l_0,l_1)$ of the ${p}^{(i)}(l_0,l_1)$.   
\begin{table}[htdp]
\caption{Estimation and 95 \% CI of $\bs{p}$.}
\label{tab:3}      
\begin{tabular}{cccccc}
\hline\noalign{\smallskip}
parameter & estimation & CI&parameter & estimation & CI \\
\noalign{\smallskip}\hline\noalign{\smallskip}
$p^{(0)}(1,1)$&$0.56060$&$[0.56055 ;0.56065]$&$p^{(1)}(1,1)$&$0.55928$&$[ 0.55923 ;  0.55933]$\\
$p^{(0)}(1,0)$&$0.03621$&$[0.03620   ; 0.03622]$&$p^{(1)}(1,0)$&$0.04707$&$[0.04706 ;0.04708 ]$\\
$p^{(0)}(0,1)$&$0.04740$&$[0.04739  ; 0.04741]$&$p^{(1)}(0,1)$&$0.03755$&$[0.03754; 0.03756 ]$\\
$p^{(0)}(0,0)$&$0.35579$&$[0.35574   ; 0.35583]$&$p^{(1)}(0,0)$&$0.35611$&$[0.35606 ; 0.35616]$\\
\noalign{\smallskip}\hline
\end{tabular}
\end{table}
The estimation of the dominant eigenvalue $\pi$ of the descendants matrix of the GW processes (characterizing extinction, see~\ref{ss:gene}) is $\wh{\pi}_9= 1.204$ with CI  $[1.191 ;  1.217 ]$. The non-extinction hypothesis ($\pi>1$) is thus satisfied.     

The means of the two reproduction laws $\bs{p}^{(0)}$ and  $\bs{p}^{(1)}$ are estimated at $\wh{m}^0_9=1.2048$ and $\wh{m}^1_9=1.2032$ respectively. The hypothesis of the equality of the mean numbers of offspring is not rejected ({p-value $=0.9$}). However, Table~\ref{tab:3} shows that there is a statistically significative difference between vectors $\bs{p}^{(0)}$ and $\bs{p}^{(1)}$ as  none of the confidence intervals intersect. Indeed, the symmetry hypothesis $\bs{p}^{(0)}=\bs{p}^{(1)}$ is rejected with {p-value $=2\cdot10^{-5}$}.
However, it is not possible to interpret this asymmetry in terms of the division of E. coli, since the cause of missing data is mostly due to observation difficulties rather than some intrinsic behavior of the cells.
%
\section{Simulation study}
\label{s:simu}
To investigate the empirical rate of convergence of our estimators as well as the power of the symmetry tests we have performed simulations of our coupled BAR-GW model. 
In particular, we study how they depend both on the ratio of missing data and on the number of observed generations.

In a complete binary tree, the number of descendants of each individual is exactly $2$. In our model of GW tree, the number of descendants is random and its average is asymptotically of the order of the dominant eigenvalue $\pi$ of the descendants matrix of the GW processes, see~\ref{ss:gene}. Therefore $\pi$ characterizes the scarcity of data: if $\pi=2$, the whole tree is observed and there are no missing data; as $\pi$ decreases, the average number of missing data increases (we choose $\pi>1$ to avoid almost sure extinction). In addition, for a single GW tree, the number of observed individuals up to generation $n$ is asymptotically of order $\pi^n$.

We have simulated the BAR-GW process for 19 distinct parameters sets, see Tables \ref{tab:setsBAR} and \ref{tab:setsGW}. Sets $1$ to $10$ are symmetric with decreasing $\pi$ (from $2$ to $1.08$), sets $11$ to $19$ are asymmetric with decreasing $\pi$ (from $1.9$ to $1.1$). The parameters of the BAR process are chosen close to the estimated values on E. coli data whereas the GW parameters are chosen to obtain different values of $\pi$. Notice that set $18$ is close to the estimated values for E. coli data. For each set, we simulated the BAR-GW process up to generation $15$ and ran our estimation procedure on $m=100$ replicated trees {($m=94$ for E. coli data)}. Each estimation was repeated $1000$ times.

\begin{table}[htp]
\begin{center}

\begin{tabular}{rlllllll}
set&$a_0$&$b_0$&$a_1$&$b_1$&$\sigma_0$&$\sigma_1$&$\rho$\\
\hline
$1$ to $10$&0.02&0.47&0.02&0.47&1.8$\cdot10^{-5}$&1.8$\cdot10^{-5}$&0.5$\cdot10^{-5}$\\
\hline
$11$ to $19$&0.0203&0.4615&0.0195&0.4782&2.28$\cdot10^{-5}$&1.34$\cdot10^{-5}$&0.48$\cdot10^{-5}$\\
\hline
\end{tabular}
\caption{Parameters sets for the simulated BAR processes.}
\label{tab:setsBAR}
\end{center}
\end{table}

\begin{table}[htp]
\begin{center}

\begin{tabular}{rccl}
set&$\bs{p}^{(0)}$&$\bs{p}^{(1)}$&$\pi$\\
\hline
$1$&(1,0,0,0)&(1,0,0,0)&2\\
$2$&(0.90,0.04,0.04,0.02)&(0.90,0.04,0.04,0.02)&1.88\\
$3$&(0.85,0.04,0.04,0.07)&(0.85,0.04,0.04,0.07)&1.78\\
$4$&(0.80,0.04,0.04,0.12)&(0.80,0.04,0.04,0.12)&1.68\\
$5$&(0.75,0.04,0.04,0.17)&(0.75,0.04,0.04,0.17)&1.58\\
$6$&(0.70,0.04,0.04,0.22)&(0.70,0.04,0.04,0.22)&1.48\\
$7$&(0.65,0.04,0.04,0.27)&(0.65,0.04,0.04,0.27)&1.38\\
$8$&(0.60,0.04,0.04,0.32)&(0.60,0.04,0.04,0.32)&1.28\\
$9$&(0.55,0.04,0.04,0.37)&(0.55,0.04,0.04,0.37)&1.18\\
$10$&(0.50,0.04,0.04,0.42)&(0.50,0.04,0.04,0.42)&1.08\\
\hline
$11$&(0.901,0.045,0.055,0.019)&(0.899,0.055,0.045,0.021)&1.9\\
$12$&(0.851,0.045,0.055,0.069)&(0.849,0.055,0.045,0.071)&1.8\\
$13$&(0.801,0.045,0.055,0.119)&(0.799,0.055,0.045,0.121)&1.7\\
$14$&(0.751,0.045,0.055,0.169)&(0.749,0.055,0.045,0.171)&1.6\\
$15$&(0.701,0.045,0.055,0.219)&(0.699,0.055,0.045,0.221)&1.5\\
$16$&(0.651,0.045,0.055,0.269)&(0.649,0.055,0.045,0.271)&1.4\\
$17$&(0.601,0.045,0.055,0.319)&(0.659,0.055,0.045,0.321)&1.3\\
$18$&(0.551,0.045,0.055,0.369)&(0.549,0.055,0.045,0.371)&1.2\\
$19$&(0.501,0.045,0.055,0.419)&(0.499,0.055,0.045,0.421)&1.1\\
\end{tabular}
\caption{Parameters sets for the simulated GW processes.}
\label{tab:setsGW}
\end{center}
\end{table}

We first investigate the significant level and power of our symmetry tests on the simulated data. The asymptotic properties of the tests are given in \ref{ss:CIBAR}. Table \ref{tab:testfixe} (resp. Table \ref{tab:testvect}) gives the proportion of reject (significant level $5\%$) under H0 (symmetric sets $1$ to $10$) and under H1 (asymmetric sets $11$ to $19$) for the test of symmetry of fixed points H0: $a_0/(1-b_0)=a_1/(1-b_1)$ (resp. the test of equality of vectors H0: $(a_0,b_0)=(a_1,b_1)$). In both cases, the proportion of reject under H0 is close to the significant level regardless of the number of observed generations (from $5$ generations on) and of the value of $\pi$. We thus can conclude that from $n=5$ on the asymptotic $\chi^2$ law is valid. Under H1, the proportion of reject increases when the number of observed generations increases and decreases when $\pi$ decreases. Recall that the number of observed individuals up to generation $n$ is asymptotically of order $m\pi^n$ ($m=100$) and the power 
is strongly linked to the number of observed data.
For instance, it is perfect for high numbers of observed generations and high $\pi$ when the expected number of observed data is huge and it is low for low $\pi$ even for high numbers of observed generations.

\begin{table}[hp]
\begin{center}

\begin{tabular}{cccccccccccc}
generation&5&6&7&8&9&10&11&12&13&14&15\\
\hline
set 1  &0.037 &0.050 &0.047 &0.048 &0.046 &0.056 &0.046 &0.047 &0.053 &0.041 &0.042\\
set 2  &0.045 &0.047 &0.047 &0.052 &0.048 &0.053 &0.050 &0.042 &0.040 &0.050 &0.049\\
set 3  &0.051 &0.048 &0.043 &0.048 &0.057 &0.064 &0.046 &0.045 &0.048 &0.049 &0.052\\
set 4  &0.051 &0.055 &0.052 &0.056 &0.049 &0.047 &0.052 &0.050 &0.059 &0.058 &0.051\\
set 5  &0.052 &0.052 &0.049 &0.053 &0.061 &0.065 &0.052 &0.054 &0.040 &0.045  &0.042\\
set 6  &0.045 &0.036 &0.039 &0.035 &0.051 &0.062 &0.054 &0.061 &0.055 &0.043  &0.046\\
set 7  &0.045 &0.048 &0.045 &0.044 &0.048 &0.037 &0.041 &0.044 &0.050 &0.049  &0.049\\
set 8  &0.046 &0.044 &0.044 &0.049 &0.047 &0.048 &0.042 &0.038 &0.043 &0.043  &0.054\\
set 9  &0.053 &0.052 &0.058 &0.061 &0.060 &0.055 &0.052 &0.052 &0.045 &0.053  &0.051\\
set 10&0.039 &0.038 &0.051 &0.046 &0.054 &0.049 &0.054 &0.046 &0.047 &0.046  &0.039\\
\hline
set 11& 0.448 &0.697 &0.926 &0.995 &1.000 &1.000 &1.000 &1.000 &1.000 &1.000 &1.000\\
set 12& 0.356 &0.568 &0.832 &0.975 &0.999 &1.000 &1.000 &1.000 &1.000 &1.000 &1.000\\
set 13& 0.305 &0.497 &0.711 &0.894 &0.991 &1.000 &1.000 &1.000 &1.000 &1.000 &1.000\\
set 14& 0.252 &0.399 &0.586 &0.777 &0.926 &0.994 &0.999 &1.000 &1.000 &1.000 &1.000\\
set 15& 0.208 &0.293 &0.417 &0.608 &0.808 &0.930 &0.990 &1.000 &1.000 &1.000 &1.000\\
set 16& 0.200 &0.279 &0.390 &0.502 &0.668 &0.790 &0.905 &0.977 &0.997 &1.000 &1.000\\
set 17& 0.174 &0.234 &0.287 &0.364 &0.458 &0.566 &0.696 &0.829 &0.912 &0.967 &0.990\\
set 18& 0.130 &0.165 &0.209 &0.255 &0.335 &0.382 &0.451 &0.548 &0.650 &0.725 &0.811\\
set 19& 0.118 &0.142 &0.174 &0.190 &0.207 &0.245 &0.300 &0.330 &0.371 &0.416 &0.459\\
\hline
\end{tabular}
\caption{Proportion of p-values $\leq 5\%$ for the equality of fixed points test (1000 replications).}
\label{tab:testfixe}
\end{center}
\end{table}

\begin{table}[hp]
\begin{center}

\begin{tabular}{cccccccccccc}
generation&5&6&7&8&9&10&11&12&13&14&15\\
\hline
set 1  &0.045    &0.062    &0.038    &0.051    &0.051    &0.051    &0.040    &0.033    &0.060    &0.036    &0.049\\
set 2  &0.036    &0.055    &0.049    &0.054    &0.044    &0.048    &0.032    &0.037    &0.039    &0.047    &0.041\\
set 3  &0.040    &0.044    &0.045    &0.053    &0.057    &0.042    &0.050    &0.039    &0.053    &0.045    &0.039\\
set 4  &0.053    &0.058    &0.055    &0.047    &0.053    &0.056    &0.061    &0.049    &0.052    &0.048    &0.043\\
set 5  &0.050    &0.050    &0.049    &0.052    &0.056    &0.049    &0.047    &0.052    &0.044    &0.048    &0.044\\
set 6  &0.058    &0.043    &0.040    &0.043    &0.052    &0.053    &0.057    &0.056    &0.048    &0.043    &0.051\\
set 7  &0.032    &0.048    &0.042    &0.032    &0.044    &0.040    &0.046    &0.035    &0.041    &0.052    &0.047\\
set 8  &0.059    &0.052    &0.058    &0.055    &0.052    &0.050    &0.053    &0.044    &0.050    &0.052    &0.050\\
set 9  &0.054    &0.049    &0.046    &0.042    &0.048    &0.042    &0.044    &0.050    &0.042    &0.047    &0.045\\
set 10&0.042    &0.049   &0.045     &0.044    &0.045    &0.053    &0.051    &0.046    &0.043    &0.044    &0.037\\
\hline
set 11& 0.414& 0.678& 0.920& 0.998& 1.000& 1.000& 1.000& 1.000& 1.000& 1.000& 1.000\\
set 12& 0.310& 0.557& 0.833& 0.980& 0.999& 1.000& 1.000& 1.000& 1.000& 1.000& 1.000\\
set 13& 0.286& 0.454& 0.703& 0.902& 0.996& 1.000& 1.000& 1.000& 1.000& 1.000& 1.000\\
set 14& 0.218& 0.367& 0.555& 0.775& 0.938& 0.995& 1.000& 1.000& 1.000& 1.000& 1.000\\
set 15& 0.193& 0.276& 0.391& 0.596& 0.789& 0.934& 0.990& 1.000& 1.000& 1.000& 1.000\\
set 16& 0.175& 0.237& 0.354& 0.479& 0.641& 0.800& 0.925& 0.980& 0.997& 1.000& 1.000\\
set 17& 0.156& 0.188& 0.246& 0.362& 0.437& 0.540& 0.683& 0.806& 0.919& 0.968& 0.989\\
set 18& 0.126& 0.152& 0.193& 0.247& 0.285& 0.359& 0.410& 0.525& 0.633& 0.726& 0.819\\
set 19& 0.110& 0.116& 0.140& 0.161& 0.192& 0.229& 0.271& 0.320& 0.365& 0.395& 0.452\\
\hline
\end{tabular}
\caption{Proportion of p-values $\leq 5\%$ for the test $(a_0,b_0)=(a_1,b_1)$ (1000 replications).}
\label{tab:testvect}
\end{center}
\end{table}

\begin{figure}[hp]
 \centerline{\includegraphics[height=8cm]{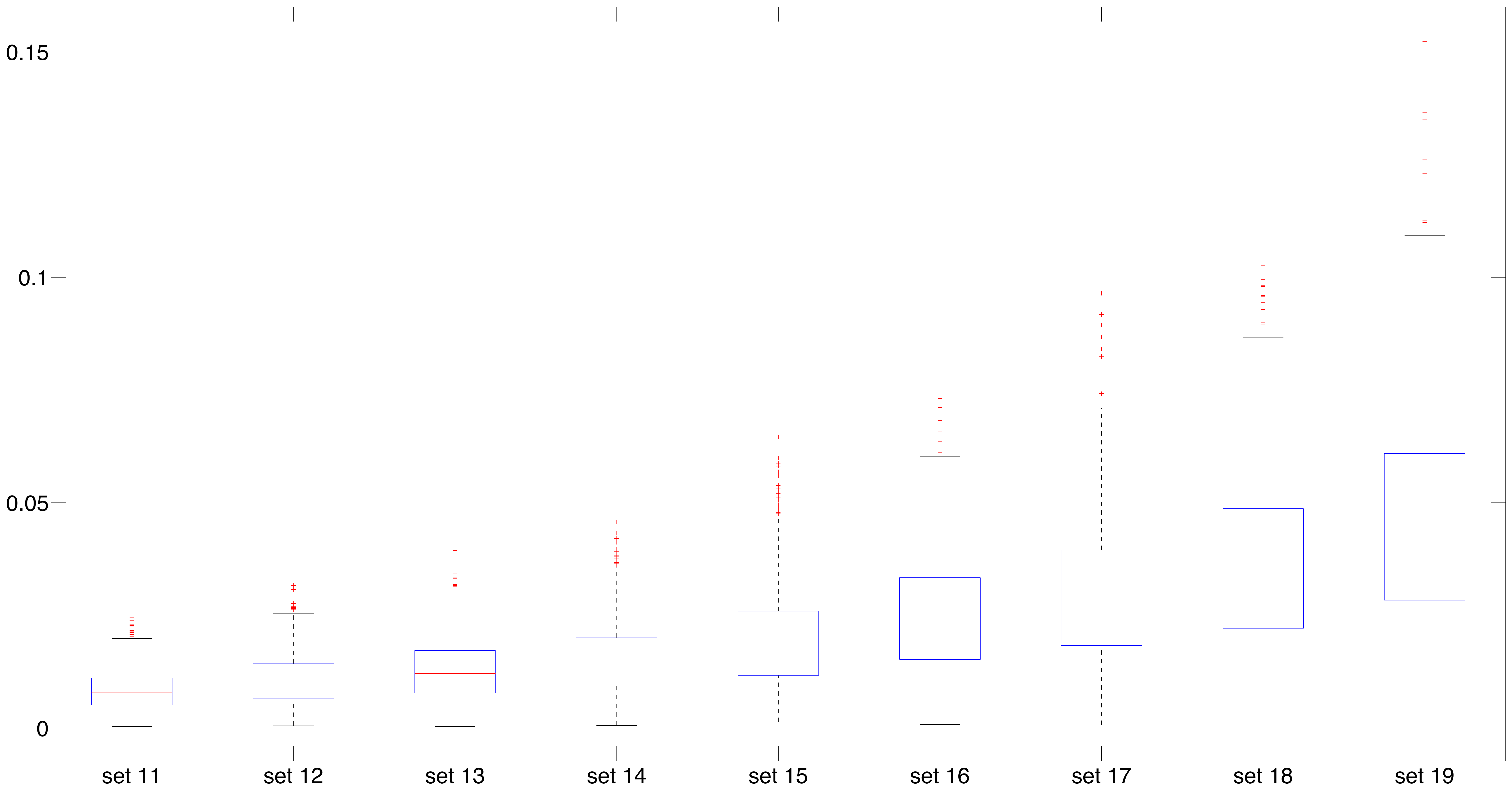}}
 
\caption{Boxplot of the estimation of the relative error $\|\wh{\bs{\theta}}_9-{\bs{\theta}}\|_2/\|{\bs{\theta}}\|_2$ for the data sets $11$ to $19$ (decreasing $\pi$)}
\label{f:BP9}
\end{figure}

\begin{figure}[hp]
 \centerline{\includegraphics[height=8cm]{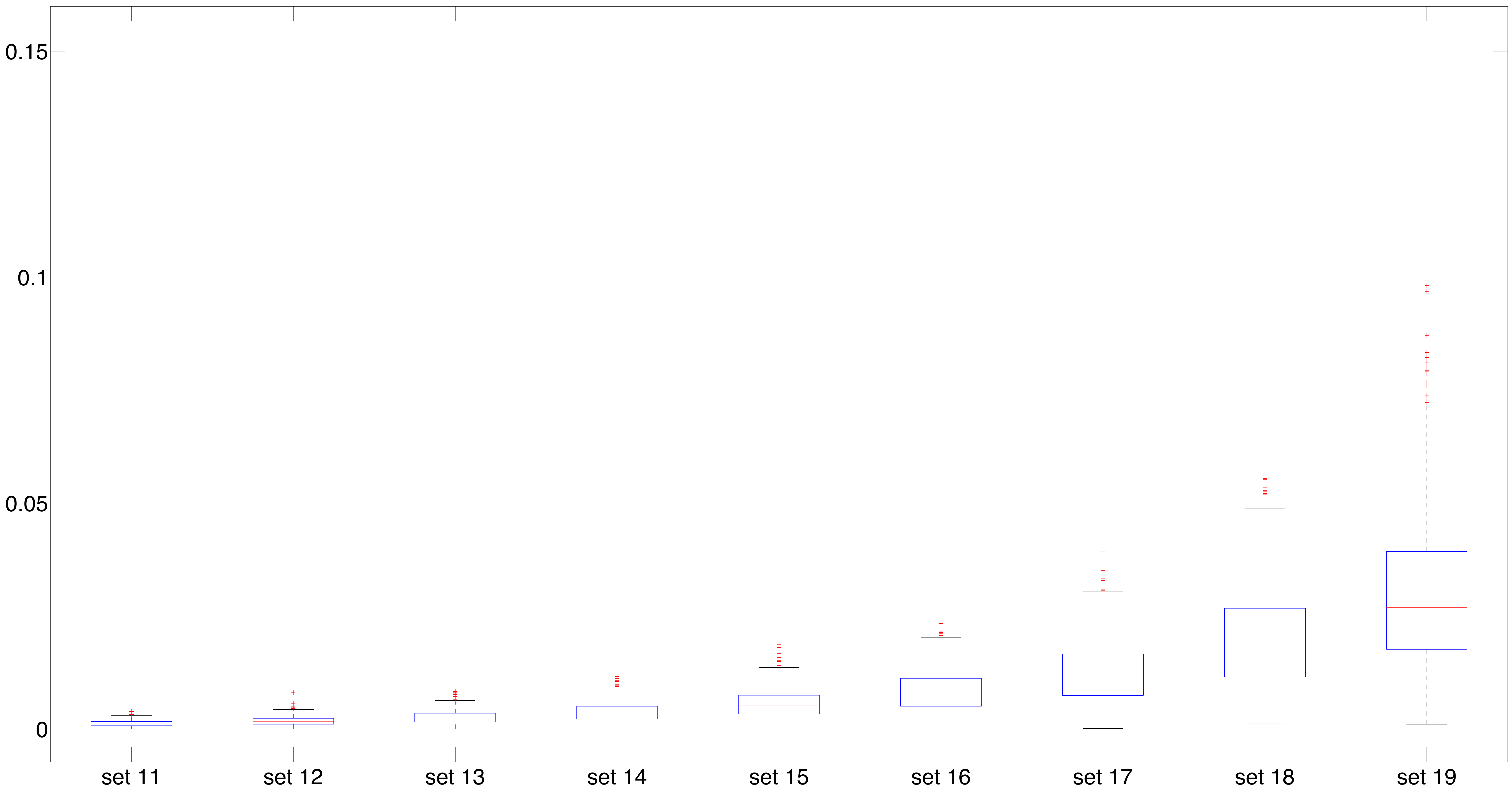}}
 
\caption{Boxplot of the estimation of the relative error $\|\wh{\bs{\theta}}_9-{\bs{\theta}}\|_2/\|{\bs{\theta}}\|_2$ for the data sets $11$ to $19$ (decreasing $\pi$)}
\label{f:BP15}
\end{figure}

Next, we investigate the empirical convergence rate of the estimation error $\|\wh{\bs{\theta}}_n-\bs{\theta}\|_2$ both as a function of the number of observed generations $n$ and of $\pi$. {Figure \ref{f:BP9} (resp. Figure \ref{f:BP15}) shows the distribution of $\|\wh{\bs{\theta}}_n-\bs{\theta}\|_2/\|{\bs{\theta}}\|_2$ for $n=9$ (reps. $n=15$) observed generations for the asymmetric parameters sets 11 to 19. It illustrates how the error deteriorates as $\pi$ decreases, i.e. as the ratio of missing data increases. The two figures have the same scale to illustrate how the relative error decreases when the number of observed generations is higher. }

\begin{figure}[h]
 \centerline{\includegraphics[height=8cm]{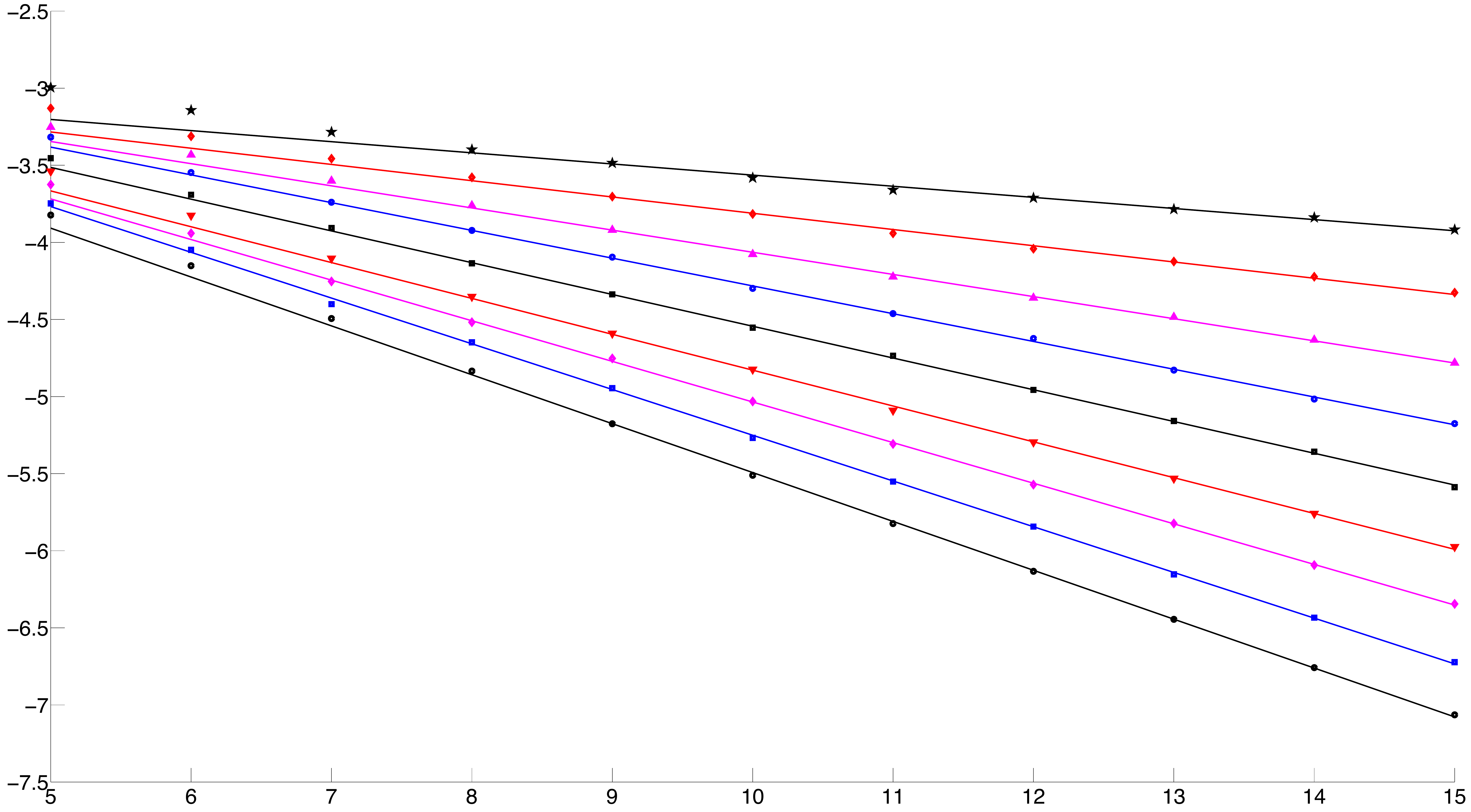}}
 
\caption{Logarithm of the error $\log\|\wh{\bs{\theta}}_{n}-{\bs{\theta}}\|_2$ as a function of the number $n$ of observed generations for the asymmetric parameters set 11 to 19 (from bottom to top: set 11-black circles, set 12-blue squares, set 13-magenta diamonds, set 14-red triangles, set 15-black squares, set 16-blue circles, set 17-magenta triangles, set 18-red diamonds, set 19-black stars).}
\label{f:cvth}
\end{figure}

{We know from Theorem \ref{th:TCL} that the variance of $\wh{\bs{\theta}}_n$ is of order $|\dT_n^*|^{-1/2}$ which asymptotically has the same order of magnitude as $\pi^{-n/2}$. In order to check how soon (in terms of the number $n$ of observed generations) this asymptotic rate is reached, we fitted the logarithm of the errors $\|\wh{\bs{\theta}}_n-\bs{\theta}\|_2$ (averaged over the 1000 simulations) to a linear function of $n$ for each parameters set (using the errors from generation {$8$} to generation $15$. The results are shown on Figure \ref{f:cvth}.}

{We also compare the computed slopes of the linear functions to the theoretical value $-\log(\pi)/2$ for the various parameters sets. The results are given in Table \ref{tab:cvrates} and show that the asymptotic rate is reached from generation $8$ on. It thus validates the accuracy of the study of E. coli data conducted in the previous section.}

\begin{table}[h]
\begin{center}

\begin{tabular}{cccccccccc}
\hspace{-2cm}set&11&12&13&14&15&16&17&18&19\\
\hline
\hspace{-2cm}empirical slope& -0.3170 &  -0.2966&   -0.2634&   -0.2325&   -0.2060&   -0.1801&-0.1413&-0.0953&-0.0672\\
\hspace{-2cm}$-log(\pi)/2$&-0.3209&-0.2939&-0.2653&-0.2350&-0.2027&-0.1682&-0.1312&-0.0912&-0.0477\\
\hline
\end{tabular}
\caption{Logarithm of empirical convergences rates vs theoretical rate}\label{tab:cvrates}
\end{center}
\end{table}
\color{black}
%
\section{Conclusion}
\label{s:conclusion}
In this paper, we first propose a statistical model to estimate and test asymmetry of a quantitative characteristic associated to each node of a family of incomplete binary trees, without aggregating single-tree estimators. An immediate application is  the investigation of asymmetry in cell lineage data. This model of coupled GW-BAR process generalizes all the previous methods on this subject in the literature because it rigorously takes into account:
\begin{itemize}
\item the dependence of the characteristic of a cell to that of its mother and the correlation between two sisters through the BAR model,
\item the possibly missing data through the GW model,
\item the information from several sets of data obtained in similar experimental conditions without the drawbacks of poor accuracy or power for small single-trees.
\end{itemize}
Furthermore, we propose the estimation of parameters of a two-type GW process in the specific context of a binary tree with a fine observation, namely the presence or absence of each cell of the complete binary tree is known. {In the context where missing offspring really come from the intrinsic reproduction, and not from faulty measures,  the asymmetry of the parameters of the GW process can be applied to cell lineage data and be interpreted as a difference in the reproduction laws between the two different types of cell.}

We applied our procedure to the E. coli data of \cite{SMT05} and concluded there exists a statistically significant asymmetry in this cell division. {Results were validated by simulation studies of the empirical rate of convergence of the estimators and power of the tests.}
 \appendix
\section{Technical assumptions and notation}
\label{s:apx1}
Our convergence results rely on martingale theory and the use of several carefully chosen filtrations regarding the BAR and/or GW process. The approach is similar to that of \citet{SGM11, SPL12}, but their results cannot be directly applied here. This is mainly due to our choice of the global non-extinction set as the union and not the intersection of the non-extinction sets of each replicated process {preventing us from directly using convergence results on single-tree estimators}. We now give some additional notation and the precise assumptions of our convergence theorems.
\subsection{Generations and extinction}
\label{ss:gene}
%
We first introduce some notation about the complete and observed genealogy trees that will be used in the sequel. For all $n\geq 1$, denote the $n$-th generation of any given tree by
$\dG_{n} = \{k,\ 2^n\leq k\leq 2^{n+1}-1\}$.
In particular, $\dG_0 = \{1\}$ is the initial generation, and $\dG_1 = \{2,3\}$ is the first generation of offspring from the first ancestor. 
Denote by 
$\dT_n = \bigcup_{\ell=0}^n\dG_{\ell}$ 
the sub-tree of all individuals from the original individual up to the $n$-th generation. Note that the cardinality $|\dG_n|$ of $\dG_n$  is $2^n$, while that of $\dT_n$ is $|\dT_n|=2^{n+1}-1$. Finally, we define the sets of observed individuals in each tree 
$\dG_{j,n}^*=\{k \in \dG_{n} : \delta_{(j,k)} =1\}$ and $\dT_{j,n}^*=\{k \in \dT_n : \delta_{(j,k)} =1\}$,
and set 
$$|\dG_n^*|= \sum_{j=1}^m |\dG_{j,n}^*|\quad \text{and}\quad  |\dT_n^*|= \sum_{j=1}^m |\dT_{j,n}^*|,$$
the total number of observed cells in all $m$ trees in generation $n$ and up to generation $n$ respectively. 
We next need to characterize the possible extinction of the GW processes, that is where $|\dT_n^*|$ does not tend to infinity with $n$. For $1\leq j\leq m$ and $n \ge 1$, we define the number of observed cells among the $n$-th generation of the $j$-th tree, distinguishing according to their type, by
$$Z_{j,n}^0=\sum_{k\in\dG_{n-1}}\delta_{(j,2k)} \quad\text{ and }\quad Z_{j,n}^1=\sum_{k\in\dG_{n-1}}\delta_{(j,2k+1)},$$
and we set $\bs{Z}_{j,n}=(Z_{j,n}^0,Z_{j,n}^1)$. For all $j$, the process $(\bs{Z}_{j,n})$ thus defined is a two-type GW process, see \cite{Har63}. We define the descendants matrix $\bs{P}$ of the GW process by
\begin{equation*}
\bs{P}=\left(\begin{array}{cc}
           p_{00} & p_{01} \\
           p_{10} & p_{11}
          \end{array}\right),
\end{equation*}
where $p_{i0} = p^{(i)}(1,0) + p^{(i)}(1,1)$ and $p_{i1} = p^{(i)}(0,1) + p^{(i)}(1,1)$, for $i \in \{0,1\}$. The quantity $p_{il}$ is thus the expected number of descendants of type $l$ of an individual of type $i$. 
It is well-known that when all the entries of the matrix $\bs{P}$ are positive, $\bs{P}$ has a positive strictly dominant eigenvalue, denoted $\pi$, which is also simple {and admits a positive left eigenvector}, see e.g.  \cite[Theorem 5.1]{Har63}. In that case, we denote by $\bs{z}=(z^0,z^1)$ the left eigenvector of  $\bs{P}$ associated with the dominant eigenvalue $\pi$ and satisfying $z^0+z^1=1$.
Let 
$\mathcal{E}_j= \bigcup_{n \ge 1} \{\bs{Z}_{j,n}=(0,0)\}$
be the event corresponding to the case when there are no cells left to observe in the $j$-th tree. We will denote $\overline{\cE}_j$ the complementary set of $\cE_j$. We are interested in asymptotic results on the set where there is an infinity of $X_{(j,k)}$ to be observed that is on the union of the non-extinction sets $\overline{\cE}_j$ denoted by 
$$\overline{\cE} = \bigcup_{j=1}^m \overline{\cE}_j=\{ \lim_{n\rightarrow \infty} |\dT_n^*| = \infty \}.$$
Note that {we allow some trees to extinct, as long as there is at least one tree still growing.} {This assumption is natural in view of the E. coli data as the collected genealogies do have a significantly different numbers of observed generations (from $4$ up to $9$).}
%
\subsection{Assumptions}
\label{ss:joint model}
Our inference is based on the $m$ i.i.d. replicas of the observed BAR process, i.e. the available information is given by the sequence $(\delta_{(j,k)}, \delta_{(j,k)}X_{(j,k)})_{1\leq j\leq m, k\geq1}$. 
We first introduce the natural generation-wise filtrations of the BAR processes. For all $1\leq j\leq m$, denote by $\dF_j=(\cF_{j,n})_{n\geq 1}$ the natural filtration associated with the $j$-th copy of the BAR process, which means that $\cF_{j,n}$ is the $\sigma$-algebra generated by all individuals of the $j$-th tree up to the $n$-th generation, $\cF_{j,n} = \sigma\{X_{(j,k)},\ k\in \dT_n\}$. For all $1\leq j\leq m$, we also define the observation filtrations as $\cO_{j,n}=\sigma\{\delta_{(j,k)}, k\in\dT_n\}$, and the sigma fields $\cO_j = \sigma \{\delta_{(j,k)} , k\geq 1 \}$. 

We make the following main assumptions on the BAR and GW processes.  
\begin{description}
\item[(H.0)] The parameters ${(a_0,b_0,a_1,b_1)}$ satisfy the usual stability assumption 
$0<\max\{|{b_0}|, |{b_1}|\} < 1$.
\item[(H.1)]  For all $1\leq j\leq m$, $n\geq 0$, $k\in \dG_{n+1}$,
$\dE[\veps_{(j,k)}^{16}]< \infty$ and $\dE[X_{(j,1)}^{16}]< \infty$.

For all $1\leq j\leq m$,  $n\geq 0$, $k\in \dG_{{n}}$ {and $i\in\{0,1\}$}, one a.s. has
\begin{equation*}
\dE[\veps_{(j,{2k+i})}|\cF_{j,n}] = 0,\ 
\dE[\veps_{(j,{2k+i})}^2|\cF_{j,n}]=\sigma^2_{{i}},\! \ 
\dE[\veps_{(j,{2k+i})}^3|\cF_{j,n}]=\lambda_{{i}},\ 
\end{equation*}
\begin{equation*}
\dE[\veps_{(j,{2k+i})}^4|\cF_{j,n}]=\tau^4_{{i}}\!,\ 
\dE[\veps_{(j,{2k+i})}^8|\cF_{j,n}]=\gamma^8_{{i}}\!,\ 
\dE[\veps_{(j,{2k+i})}^{16}|\cF_{j,n}]=\mu^{16}_{{i}}\!.
\end{equation*}
For all $1\leq j\leq m$, $n\geq 0$, $k \in \dG_n$, one a.s. has
\begin{equation*}
\dE[\veps_{(j,2k)}\veps_{(j,2k+1)}|\cF_{j,n}] = \rho,\quad \dE[\veps_{(j,2k)}^2\veps_{(j,2k+1)}^2|\cF_{j,n}] = \nu^2\!,\quad 
\dE[\veps_{(j,2k)}^8\veps_{(j,2k+1)}^8|\cF_{j,n}] = \eta^8\!, 
\end{equation*}\begin{equation*}
\dE[\veps_{(j,2k)}^2\veps_{(j,2k+1)}|\cF_{j,n}] = \alpha, \quad
\dE[\veps_{(j,2k)}\veps_{(j,2k+1)}^2|\cF_{j,n}] = \beta.\end{equation*}
\item[(H.2)] For all $1\leq j\leq m$ and $n\geq 0$ the vectors $\{(\veps_{(j,2k)} , \veps_{(j,2k+1)}),\ k \in\dG_n\}$ are conditionally independent given~$\cF_{j,n}$.
\item[(H.3)] The sequences $(\veps_{(1,k)})_{k\geq 2}, (\veps_{(2,k)})_{k\geq 2}, \ldots , (\veps_{(m,k)})_{k\geq 2}$ are independent. The random variables  $(X_{(j,1)})_{1\leq j\leq m}$ are independent and independent from the noise sequences.
 \item[(H.4)] For all $1\leq j\leq m$, the sequence $(\delta_{(j,k)})_{k\geq 1}$ is independent from the sequences $(X_{(j,k)})_{k\geq 1}$ and $(\veps_{(j,k)})_{k\geq 2}$.
\item[(H.5)] The sequences $(\delta_{(1,k)})_{k\geq 2}, (\delta_{(2,k)})_{k\geq 2}, \ldots , (\delta_{(m,k)})_{k\geq 2}$ are independent.
\end{description}
We also make the following super criticality assumption on the matrix $\bs{P}$.
\begin{description}
 \item[(H.6)] All entries of the matrix $\bs{P}$ are positive: for all $(i,l) \in \{0,1\}^2$, $p_{il} > 0$, and the dominant eigenvalue is greater than one: $\pi > 1$ .
\end{description}
If $\pi>1$, it is well known, see e.g. \cite{Har63}, that the extinction probability of the GW processes is less than one: for all $1\leq j\leq m$, $\mathbb{P}(\cE_j)=p<1$. Under assumptions {(H.5-6)}, one thus clearly has
$\mathbb{P}(\overline{\cE})=1-p^m>0.$ 

{Note that under these assumptions, it is proved in \cite{SGM11} that the single-tree estimators $\wh{\bs{\theta}}_{j,n}$ are consistent on the single-tree non-extinction sets $\overline{\cE}_j$. This result is based on the separate convergence of $\bs{\Sigma}_{j,n}$ and $\bs{\Sigma}_{j,n}\wh{\bs{\theta}}_{j,n}$. Therefore, the convergence of our global estimator $\wh{\bs{\theta}}_{n}$ is readily obtained on the intersection of the single-tree non-extinction sets $\cap_{j=1}^m\overline{\cE}_j$, see Section \ref{s:methodo}. However, we are interested in the convergence of the global estimator on the larger set $\overline{\cE}=\cup_{j=1}^m\overline{\cE}_j$. This is why we cannot directly use the results of \cite{SGM11}. We explain in the following sections how the ideas therein have to be adapted to this new framework.}

%
\subsection{Additional estimators}
\label{ss:estimplus}
From the estimators of the reproductions probabilities of the GW process, one can easily construct an estimator of the spectral radius $\pi$ of the descendants matrix $\bs{P}$ of the GW process. Indeed, $\bs{P}$ is a $2\times2$ matrix so that its spectral radius can be computed explicitly as a function of its coefficients, namely
\begin{equation*}
\pi=\frac{1}{2}\Big(tr(\bs{P})+\big(tr(\bs{P})^2-4\det(\bs{P})\big)^{1/2}\Big).
\end{equation*}
Replacing the coefficients of $\bs{P}$ by their empirical estimators, one obtains
\begin{equation*}
\wh{\pi}_n = \frac{1}{2}\big(\wh{T}_n+(\wh{T}_n^2-4\wh{D}_n)^{1/2}\big).
\end{equation*}
where 
\begin{eqnarray*}
\wh{T}_n&=&\wh{p}_n^{(0)}(1,0)+\wh{p}_n^{(0)}(1,1)+\wh{p}_n^{(1)}(0,1)+\wh{p}_n^{(1)}(1,1),\\
\wh{D}_n&=&(\wh{p}_n^{(0)}(1,0)+\wh{p}_n^{(0)}(1,1))(\wh{p}_n^{(1)}(0,1)+\wh{p}_n^{(1)}(1,1))-(\wh{p}_n^{(0)}(0,1)+\wh{p}_n^{(0)}(1,1))(\wh{p}_n^{(1)}(1,0)+\wh{p}_n^{(1)}(1,1))
\end{eqnarray*}
are the empirical estimator of the trace $tr(\bs{P})$ and the determinant $\det(\bs{P})$ respectively.
Finally, to compute confidence intervals for $\sigma^2_{{i}}$ and $\rho$, we need an estimation of higher moments. We use again empirical estimators
$$\wh{\tau}^4_{{i},n} = \frac{1}{ |\dT_{n-1}^{*{i}}|} \sum_{j=1}^m \sum_{k \in
\dT_{n-1}} \wh{\veps}_{(j,2k+{i})}^4,
\qquad 
\wh{\nu}_{n}^2 = \frac{1}{|\dT_{n-1}^{*01}|} \sum_{j=1}^m \sum_{k \in\dT_{n-1}} \wh{\veps}_{(j,2k)}^2 \wh{\veps}_{(j,2k+1)}^2.$$ 
%
\section{Convergence of estimators for the GW process}
\label{ss:apxGW}
We now prove the convergence of the estimators for the GW process, that is the first parts of Theorems~\ref{th:CVp} and \ref{th:TCL}, together with additional technical results.
\subsection{Preliminary results: from single-trees to multiple trees}
\label{ss:apxGWprelim}
Our objective is  to show that we can adapt the results in \cite{SGM11} to the multiple tree framework {despite our choice of considering the union and not the intersection of the single-tree non-extinction sets}. To this aim, we first need to recall Lemma~A.3 of \cite{BSG09}. 

\begin{lemma}
\label{lemtoepbar}
Let $(\bs{A}_n)$ be a sequence of real-valued matrices such that
\begin{equation*}
 \sum_{n=0}^{\infty}\|\bs{A}_n\|<\infty\qquad\textrm{and}\qquad\lim_{n \rightarrow \infty} \sum_{k=0}^n\bs{A}_k=\bs{A}.
\end{equation*}
In addition, let $(\bs{X}_n)$ be a sequence of real-valued vectors which converges to a limiting value $\bs{X}$.
Then, one has
\begin{equation*}
\lim_{n \rightarrow \infty}{\sum_{\ell=0}^n \bs{A}_{n-\ell}\bs{X}_{\ell}}=\bs{AX}.
\end{equation*}
\end{lemma}

The next result is an adaptation of Lemma A.2 in \cite{BSG09} to the GW tree framework. It gives a correspondence between sums on one generation and sums on the whole tree.
\begin{lemma}
\label{lemTG}
Let $(x_n)$ be a sequence of real numbers and $\pi>1$. One has 
\begin{equation*}
\lim_{n\rightarrow\infty}\frac{1}{\pi^n}\sum_{k\in\dT_n}x_k=x\Longleftrightarrow\lim_{n\rightarrow\infty}\frac{1}{\pi^n}\sum_{k\in\dG_n}x_k=\frac{\pi-1}{\pi}x.
\end{equation*}
\end{lemma}

\noindent\textit{Proof:} Suppose that $\pi^{-n}\sum_{k\in\dT_n}x_k$ converges to $x$. Then one has
\begin{equation*}
\frac{1}{\pi^n}\sum_{k\in\dG_n}x_k=\frac{1}{\pi^n}\sum_{k\in\dT_n}x_k-\frac{1}{\pi}\frac{1}{\pi^{n-1}}\sum_{k\in\dT_{n-1}}x_k
\xrightarrow[{n\rightarrow\infty}]{}x-\frac{1}{\pi}x\ =\ \frac{\pi-1}{\pi}x.
\end{equation*}
Conversely, if $\pi^{-n}\sum_{k\in\dG_n}x_k$ converges to $y$, as $\dT_n=\cup_{\ell=0}^n\dG_{\ell}$, one has
\begin{equation*}
\frac{1}{\pi^n}\sum_{k\in\dT_n}x_k=\sum_{\ell=0}^n\frac{1}{\pi^{n-\ell}}\frac{1}{\pi^{\ell}}\sum_{k\in\dG_{\ell}}x_k
\xrightarrow[{n\rightarrow\infty}]{}\frac{\pi}{\pi-1}y,
\end{equation*}
using Lemma~\ref{lemtoepbar} with $\bs{A}_n=\pi^{-n}$ and $\bs{X}_{n}=\pi^{-n}\sum_{k\in\dG_n}x_k$.
\hspace{\stretch{1}}$ \Box$\\

We now adapt Lemma~2.1 of \cite{SGM11} to our multiple tree framework.

\begin{lemma}
\label{pi-T} 
Under assumption \emph{ (H.5-6)}, there exist a nonnegative random variable $W$ such that for all sequences $(x_{(1,n)}),\ldots, (x_{(m,n)})$ of real numbers one has a.s.
\begin{equation*}
\lim_{n\rightarrow\infty}\frac{\ind{|\dG_{n}^*|>0}}{|\dT^*_{n}|}\sum_{j=1}^m\sum_{k\in\dT_n}x_{(j,k)}=x\indnex\Longleftrightarrow
\lim_{n\rightarrow\infty}\frac{1}{\pi^n}\sum_{j=1}^m\sum_{k\in\dT_n}x_{(j,k)}=x\frac{\pi}{\pi-1}W.
\end{equation*}
\end{lemma}

\noindent\textit{Proof:} We use a well known property of super-critical GW processes, see e.g. \cite{Har63}: for all $j$, there exists a non negative random variable $W_j$ such that
\begin{equation}\label{lim T star}
 \lim_{n\rightarrow \infty} \frac{|\dT^*_{j,n}|} {\pi^n}= \frac{\pi}{\pi-1} W_j    \quad \text{a.s.}
\end{equation}
and in addition $\{W_j>0\}=\overline{\cE}_j=\lim\{|\dG_{j,n}^*|>0\}$. Therefore, one has
\begin{equation*}
\lim_{n\rightarrow \infty}\sum_{j=1}^m\frac{|\dT^*_{j,n}|} {\pi^n}=\lim_{n\rightarrow \infty}\frac{|\dT^*_{n}|} {\pi^n}=\frac{\pi}{\pi-1} \sum_{j=1}^mW_j  \quad \text{a.s.}
\end{equation*}
The result is obtained by setting $W=\sum_{j=1}^mW_j$ and noticing that $\overline{\cE}=\cup_{j=1}^m\overline{\cE}_j=\{\sum_{j=1}^mW_j>0\}=\lim\{|\dG_{n}^*|>0\}$.
\hspace{\stretch{1}}$ \Box$\\

Finally, the main result of this section is new and explains how convergence results on multiple trees can be obtained from convergence results on a single-tree. This will allow us to directly use results from \cite{SGM11} in all the sequel.

\begin{lemma}
\label{lemLLNtree}
Let $(x_{(1,n)}),\ldots, (x_{(m,n)})$ be $m$ sequences of real numbers such that for all $1\leq j\leq m$ one has the a.s. limit
\begin{equation}\label{lim tree}
\lim_{n\rightarrow \infty} \frac{\ind{|\dG_{j,n}^*|>0}}{|\dT^*_{j,n}|}\sum_{k\in\dT_n}x_{(j,k)}=  \ell \indnexj,
\end{equation}
then under assumptions \emph{ (H.5-6)} one also has
\begin{equation*}
\lim_{n\rightarrow \infty} \frac{\ind{|\dG_{n}^*|>0}}{|\dT^*_{n}|}\sum_{j=1}^m\sum_{k\in\dT_n}x_{(j,k)}=  \ell \indnex\quad \text{a.s.}
\end{equation*}
\end{lemma}

\noindent\textit{Proof:} Equations~\reff{lim tree} and \reff{lim T star} yield, for all $j$,
\begin{equation*}
\lim_{n\rightarrow \infty} \frac{1}{\pi^n}\sum_{k\in\dT_n}x_{(j,k)}=  \ell \frac{\pi-1}{\pi} W_j.
\end{equation*}
Summing over $j$, one obtains
\begin{equation*}
\lim_{n\rightarrow \infty} \frac{1}{\pi^n}\sum_{j=1}^m\sum_{k\in\dT_n}x_{(j,k)}=  \ell \frac{\pi-1}{\pi} \sum_{j=1}^mW_j=\ell \frac{\pi-1}{\pi}W.
\end{equation*}
Finally, we use Lemma~\ref{pi-T} to conclude.
\hspace{\stretch{1}}$ \Box$
%
\subsection{Strong consistency for the estimators of the GW process}
\label{ss:apxGWconsist}
To prove the convergence of the $\wh{p}_n^{(i)}(l_0,l_1)$ we first need to derive a convergence result for a sum of independent GW processes.

\begin{lemma}
\label{lemmaGWapx}
Suppose that assumptions~\emph{({H.5-6})} are satisfied. Then for $i\in\{0,1\}$ one has 
$$\lim_{n\rightarrow \infty} \ind{|\dG_n^*|>0}{|\dT_n^*|}^{-1}\sum_{j=1}^m\sum_{k\in\dT_{n-1}}\delta_{(j,2k+i)} = z^i \indnex\quad\text{a.s.}$$ 
\end{lemma}

\noindent\textit{Proof } 
Remarking that $\sum_{j=1}^m\sum_{k\in\dT_{n-1}}\delta_{(j,2k+i)} = \sum_{j=1}^m\sum_{l=1}^n{Z}_{j,l}^i$,  
the lemma is a direct consequence of Lemma~\ref{lemLLNtree} and the well-known property of super-critical GW processes $\ind{|\dG_{j,n}^*|>0}|\dT^*_{j,n}|^{-1}\sum_{l=1}^n\bs{Z}_{j,l} \rightarrow  \bs{z} \indnex$, for all $0\leq j \leq m$.
\hspace{\stretch{1}}$ \Box\\$

\noindent\textit{Proof of Theorem~\ref{th:CVp}, first part} 
We give the details of the convergence of $\wh{p}_n^{(1)}(1,1)$ to $p^{(1)}(1,1)$, the other convergences are derived similarly. The proof relies on the convergence of square integrable scalar martingales.
Set
\begin{equation*}
M_n={\sum_{j=1}^m\sum_{k\in\dT_{n-2}}\delta_{(j,2k+1)}\big(\delta_{(j,4k+2)}\delta_{(j,4k+3)}-p^{(1)}(1,1)\big)}.
\end{equation*}
We are going to prove that $(M_n)$ is a martingale for a well chosen filtration.
Recall that $\cO_{j,n}=\sigma\{\delta_{(j,k)}, k\in\dT_n\}$,  and set $\cO_{n}=\vee_{j=1}^m\cO_{j,n}$. Then $(M_n)$ is clearly a square integrable real $(\cO_{n})$-martingale. Using the independence assumption (H.5), its increasing process is
\begin{equation*}
<M>_n=\sum_{j=1}^m\sum_{k\in\dT_{n-2}}\delta_{(j,2k+1)}p^{(1)}(1,1)\big(1-p^{(1)}(1,1)\big)=p^{(1)}(1,1)\big(1-p^{(1)}(1,1)\big)\sum_{j=1}^m\sum_{\ell=0}^{n-1}Z_{j,\ell}^1.
\end{equation*}
Hence, Lemma~\ref{lemmaGWapx} implies that $|\dT_{n-1}^*|^{-1}<M>_n$ converges almost surely on the non-extinction set $\overline{\cE}$. The law of large numbers for scalar martingales thus yields that $|\dT_{n-1}^*|^{-1}M_n$ tends to $0$ as $n$ tends to infinity on $\overline{\cE}$. Finally, notice that
\begin{equation*}
\wh{p}_n^{(1)}(1,1)-p^{(1)}(1,1)=\frac{M_n}{\sum_{j=1}^m\sum_{k\in\dT_{n-2}}\delta_{(j,2k+1)}}=\frac{M_n}{\sum_{j=1}^m\sum_{\ell=0}^{n-1}Z_{j,\ell}^1},
\end{equation*}
so that Lemma~\ref{lemmaGWapx} again implies the almost sure convergence of $\wh{p}_n^{(1)}(1,1)$ to $p^{(1)}(1,1)$ on the non-extinction set $\overline{\cE}$.
\hfill$\Box$\\

As a direct consequence, one obtains the a.s. convergence of $\wh{\pi}_n$ to $\pi$ on $\overline{\cE}$.
%
\subsection{Asymptotic normality for the estimators of the GW process}
\label{ss:apxGWnorm}
As $\dP(\overline{\cE})\neq 0$, we can define a new probability  $ \dP_{\overline{\cE}}$ by $\dP_{\overline{\cE}}(A) =  {\dP(A \cap \overline{\cE})}/{ \dP(\overline{\cE})}$ for all event $A$. In all the sequel of this section, we will work on the space $\overline{\cE}$ under the probability $\mathbb{P}_{\overline{\cE}}$ and we denote by $\mathbb{E}_{\overline{\cE}}$ the corresponding expectation.
 We can now turn to the proof of the asymptotic normality of $\wh{\bs{p}}_n$. The proof also relies on martingale theory. As the normalizing term in our central limit theorem is random, we use the central limit theorem for martingales given in Theorem~2.1.9 of \citet{Duflo97} that we first recall as Theorem~\ref{TCL:Dufloapx} for self-completeness.

\begin{theorem}\label{TCL:Dufloapx}
Suppose that $(\Omega, \cA, P)$ is a probability space and that for each $n$ we have a filtration $\dF_n=(\cF_k^{(n)})$, a stopping time $\nu_n$ relative to $\dF_n$ and a real, square-integrable vector martingale $M^{(n)}=(M_k^{(n)})_{k\geq0}$ which is adapted to $\dF_n$ and has hook denoted by $<M>^{(n)}$. We make the following two assumptions.
\begin{description}
\item[A.1] For a deterministic symmetric positive semi-definite matrix $\Gamma$
$$<M>^{(n)}_{\nu_n}\xrightarrow[]{P}\Gamma.$$
\item[A.2] Lindeberg's condition holds; in other words, for all $\epsilon>0$,
$$\sum_{k=1}^{\nu_n}\dE\Big(\|M_k^{(n)}-M_{k-1}^{(n)}\|^2\ind{\|M_k^{(n)}-M_{k-1}^{(n)}\|>\epsilon}\ |\ \cF_{k-1}^{(n)}\Big)\xrightarrow[]{P}0.$$
\end{description}
Then:
$$M^{(n)}_{\nu_n}\xrightarrow[]{\cL}\cN(0,\Gamma).$$
\end{theorem}

\noindent\textit{Proof of Theoremn\ref{th:TCL}, first part} 
First, set 
\begin{equation}\label{defV}
\bs{V}=\left(\begin{array}{cc}
\bs{V}^0/z^0&0\\
0&\bs{V}^1/{z^1}
\end{array}\right)
\end{equation}
where for all $i$ in $\{0,1\}$, $\bs{V}^i=\bs{W}^i-\bs{p}^{(i)}(\bs{p}^{(i)})^t$, $\bs{W}^i$ is a $4\times 4$ matrix with the entries of $\bs{p}^{(i)}$ on the diagonal and $0$ elsewhere.
We are going to prove that $\bs{V}$ is the asymptotic variance of $\wh{\bs{p}}_n-\bs{p}$ suitably normalized.
We use Theorem~\ref{TCL:Dufloapx}. We first need to define a suitable filtration. Here, we use the first cousins filtration defined as follows. Let 
$$\mathcal{H}_{j,p}=\sigma\{\delta_{(j,1)},\ldots \delta_{(j,3)}, (\delta_{(j,4k)}, \ldots, \delta_{(j,4k+3)}),1\leq k\leq p\}$$ 
be the $\sigma$-field generated by all the $4$-tuples of observed cousin cells up the granddaughters of cell $(j,p)$ in the $j$-th tree and 
$\mathcal{H}_{p}=\vee_{j=1}^m \mathcal{H}_{j,p}$.
Hence,  the $4$-tuple $(\delta_{(j,4k)}, \ldots, \delta_{(j,4k+3)})$ is $\mathcal{H}_k$-measurable for all $j$.
By definition of the reproduction probabilities $p^{(i)}(l_0,l_1)$, the processes
$$\big(\delta_{(j,2k+i)}(\phi_{l_0}(\delta_{(j,2(2k+i))})\phi_{l_1}(\delta_{(j,2(2k+i)+1)})- p^{(i)}(l_0,l_1)\big)_{k\geq 1}$$ 
are $(\mathcal{H}_k)$-martingale difference sequences. We thus introduce a sequence of $(\mathcal{H}_k)$-mar\-tin\-ga\-les $(\bs{M}^{(n)}_p)_{\{p \geq 1\}} $ defined for all $n\geq 1$ and $p\geq 1$ by 
$$\bs{M}^{(n)}_p =   {|\dT^*_{n-1}|}^{-1/2} \sum_{k=1}^{p}\sum_{j=1}^m\bs{D}_{(j,k)},$$
with
$\bs{D}_{(j,k)}=\big((\bs{D}_{(j,k)}^0)^t, (\bs{D}_{(j,k)}^1)^t\big)^t$ and

\begin{equation*}
\bs{D}_{(j,k)}^i=\delta_{(j,2k+i)}\left( \begin{array}{c}
 \delta_{(j,2(2k+i))}\delta_{(j,2(2k+i)+1)} - p^{(i)}(1,1) \\
 \delta_{(j,2(2k+i))}(1-\delta_{(j,2(2k+i)+1)}) - p^{(i)}(1,0)\\
( 1-\delta_{(j,2(2k+i))})\delta_{(j,2(2k+i)+1)} - p^{(i)}(0,1)\\
( 1-\delta_{(j,2(2k+i))})(1-\delta_{(j,2(2k+i)+1)}) - p^{(i)}(0,0)
\end{array}\right).
\end{equation*}
\color{black}
We also introduce the sequence of stopping times $\nu_n=|\dT_{n-2}|=2^{n-1}-1$. One has
\begin{equation*}
\mathbb{E}_{\overline{\cE}}[\bs{D}_{(j,k)}\bs{D}_{(j,k)}^t|\mathcal{H}_{k-1}]= 
\left(
\begin{array}{cc}
\delta_{(j,2k)}\bs{V}^0&0\\
0& \delta_{(j,2k+1)}\bs{V}^1
\end{array}\right).
\end{equation*} 
Therefore the one has $<\bs{M}^{(n)}>_{\nu_n}=|\dT_{n-1}^*|^{-1}\sum_{j=1}^m\sum_{\ell=0}^{n-1}\left(
\begin{array}{cc}
Z_{j,\ell}^0\bs{V}^0&0\\
0&Z_{j,\ell}^1\bs{V}^1
\end{array}\right)$, so that its $\mathbb{P}_{\overline{\cE}}$ almost sure limit is
\begin{equation*}
\bs{\Gamma}'=\left(\begin{array}{cc} z^0\bs{V}^0&0\\
0&z^1\bs{V}^1
\end{array}\right),
\end{equation*}
thanks to Lemma~\ref{lemmaGWapx}. 
Therefore, assumption A.1 of Theorem~\ref{TCL:Dufloapx} holds under
$\mathbb{P}_{\overline{\cE}}$.  The Lindeberg condition A.2 is obviously satisfied as we deal with {finite support distributions}. We then conclude that under
$\mathbb{P}_{\overline{\cE}}$ one has 
$${{|\dT^*_{n-1}|}}^{-1/2}\bs{M}^{(n)}_{\nu_n}={{|\dT^*_{n-1}|}}^{-1/2}\sum_{j=1}^m\sum_{k \in \mathbb{T}^*_{n-2}}\bs{D}_{(j,k)}
\xrightarrow{\cL}
\cN(0,\bs{\Gamma'}).$$ 
Using the relation 
$$\wh{\bs{p}}_{n}- \bs{p} = \left(
\begin{array}{cc}
(\sum_{j=1}^m\sum_{\ell=0}^{n-1} Z_{j,\ell}^0)\rI_4&0\\
0&(\sum_{j=1}^m\sum_{\ell=0}^{n-1} Z_{j,\ell}^1)\rI_4
\end{array}\right) ^{-1} \bs{M}^{(n)}_{\nu_n},$$
Lemma~\ref{lemmaGWapx} and Slutsky's Lemma give the first part of Theorem \ref{th:TCL}.
\hfill$\Box$
%
\subsection{Interval estimation and tests for the GW process}
\label{ss:CIGW}
From the central limit theorem~\ref{th:TCL} one can easily build asymptotic confidence intervals for our estimators. In our context, $Y_n$ and $Y'_n$ being two random variables, we will say that $[Y_n ; Y_n']$ is an asymptotic confidence interval with confidence level $1 - \epsilon$ for the parameter $Y$ if 
$\dP_{\overline{\cE}}\big(Y_n \leq Y \leq Y_n'\big)\xrightarrow[n\rightarrow\infty]{}(1-\epsilon).$
For any $0\leq \epsilon\leq 1$, let $q_{1-\epsilon/2}$ be the $1-\epsilon/2$ quantile of the standard normal law.

For all $n\geq 2$, define the $8\times8$ matrix
$$\wh{\bs{V}}_n=\left(
\begin{array}{cc}
\wh{\bs{V}}_n^0\big(\sum_{j=1}^m\sum_{k\in\dT_{n-2}}\delta_{(j,2k)}\big)^{-1}&0\\
0&\wh{\bs{V}}_n^1\big(\sum_{j=1}^m\sum_{k\in\dT_{n-2}}\delta_{(j,2k+1)}\big)^{-1},
\end{array}\right)$$
where for all $i$ in $\{0,1\}$, $\wh{\bs{V}}^i_n=\wh{\bs{W}}^i_n-\wh{\bs{p}}^{(i)}_n(\wh{\bs{p}}^{(i)}_n)^t$, $\wh{\bs{W}}^i_n$ is a $4\times 4$ matrix with the entries of $\wh{\bs{p}}^{(i)}_n$ on the diagonal and $0$ elsewhere. Thus, $|\dT_{n-1}^*|\wh{\bs{V}}_n$ is an empirical estimator of the covariance matrix $\bs{V}$.
\begin{theorem}
\label{th:ICpapx}
Under assumptions \emph{({H.5-6})}, for $i,l_0,l_1$ in $\{0,1\}$ and for any $0< \epsilon< 1$, the  random interval defined by $$
\big[\wh{p}_n^{(i)}(l_0,l_1)-q_{1-\epsilon/2}{(\wh{\bs{V}}_{n}^{1/2})_{\ell,\ell}} \;\;; \;\;\wh{p}_n^{(i)}(l_0,l_1)+q_{1-\epsilon/2}(\wh{\bs{V}}_{n}^{1/2})_{\ell,\ell}\big]
$$is an asymptotic confidence interval with level $1 - \epsilon$ for $p^{(i)}(l_0,l_1) $;
where $(\ell,\ell)$ is the coordinate of $\bs{V}_{n}$ corresponding to $p^{(i)}(l_0,l_1)$, namely {$\ell=4(i+1)-(2l_0+l_1)$}.
\end{theorem}

\noindent\textit{Proof} This is a straightforward consequence of the central limit Theorem~\ref{th:TCL} together with Slutsky's lemma as 
$
\lim_{n\rightarrow\infty}|\dT_{n-1}^*|\wh{\bs{V}}_{n}=\bs{V}
$ 
$\mathbb{P}_{\overline{\cE}}$ a.s. thanks to Lemma~\ref{lemmaGWapx} and Theorem~\ref{th:CVp}.
\hspace{\stretch{1}}$ \Box$\\

Set $\wh{{G}}_n=\wh{\bs{F}}_n^t\wh{\bs{V}}_{n}\wh{\bs{F}}_n$, where $\wh{\bs{F}}_n$ is the $8\times1$ vector defined by

\begin{equation*}
\wh{\bs{F}}_n=\frac{1}{2}{\left(1,1,0,0,1,0,1,0\right)}^t+\frac{1}{2}(\wh{T}_n^2-4\wh{D}_n)^{-1/2}\wh{\bs{H}}_n
\end{equation*}
and

\begin{equation*}
\wh{\bs{H}}_n=\left(\begin{array}{c}
\wh{p}_n^{(0)}(1,1)+\wh{p}_n^{(0)}(1,0)+\wh{p}_n^{(1)}(1,1)+2\wh{p}_n^{(1)}(1,0)-\wh{p}_n^{(1)}(0,1)\\
\wh{p}_n^{(0)}(1,1)+\wh{p}_n^{(0)}(1,0)-\wh{p}_n^{(1)}(1,1)-\wh{p}_n^{(1)}(0,1)\\
2\wh{p}_n^{(1)}(1,1)+2\wh{p}_n^{(1)}(1,0)\\
0\\
\wh{p}_n^{(0)}(1,1)-\wh{p}_n^{(0)}(1,0)+2\wh{p}_n^{(0)}(0,1)+\wh{p}_n^{(1)}(1,1)+\wh{p}_n^{(1)}(0,1)\\
2\wh{p}_n^{(0)}(1,1)+2\wh{p}_n^{(0)}(0,1)\\
-\wh{p}_n^{(0)}(1,1)-\wh{p}_n^{(0)}(1,0)+\wh{p}_n^{(1)}(1,1)+\wh{p}_n^{(1)}(0,1)\\
0
\end{array}\right).
\end{equation*}
\color{black}
\begin{theorem}
\label{th:ICpiapx}
Under assumptions \emph{({H.5-6})}, for any $0< \epsilon< 1$ one has that
\begin{equation*}
\big[\wh{\pi}_n-q_{1-\epsilon/2}{\wh{{G}}_{n}^{1/2}} \;\; ; \;\; \wh{\pi}_n+q_{1-\epsilon/2}\wh{{G}}_{n}^{1/2}\big]
\end{equation*}is an asymptotic confidence interval with level $1 - \epsilon$ for  $ \pi $.
\end{theorem}

\noindent\textit{Proof} This is again a straightforward consequence of the central limit Theorem~\ref{th:TCL} together with Slutsky's lemma as $\wh{\bs{F}}_n$ is the gradient of the function that maps the vector $\wh{\bs{p}}$ onto the estimator $\wh{\pi}_n$.
\hspace{\stretch{1}}$ \Box$\\

We propose two symmetry tests for the GW process. The first one compares the average number of offspring $m_0$ of a cell of type $0$: $m_0=p^{(0)}(1,0)+p^{(0)}(0,1)+2p^{(0)}(1,1)$ to that of a cell of type $1$:  $m_1 = p^{(1)}(1,0)+p^{(1)}(0,1)+2p^{(1)}(1,1)$. Denote by $\wh{m}_n^0$ and $\wh{m}_n^1$ their empirical estimators.
Set
\begin{itemize}
\item $\bs{H_0^{m}}$: $m_0=m_1$ the symmetry hypothesis,
\item $\bs{H_1^{m}}$: $m_0 \neq m_1$ the alternative hypothesis.
\end{itemize}
Let $Y_n^{m}$ be the test statistic defined by
\begin{equation*}
Y_n^{m}=|\dT_{n-1}^*|^{1/2}(\wh{\Delta}_n^{m})^{-1/2}(\wh{m}_n^0-\wh{m}_n^1),
\end{equation*}
where
$\wh{\Delta}_n^{m} = |\dT_{n-1}^*|\bs{dg}_{m}^t\wh{\bs{V}}_n\bs{dg}_{m}$ and $\bs{dg}_{m}={(2,1,1,0,-2 -1,-1,0)}^t$. 
This test statistic has the following asymptotic properties.
\begin{theorem}
\label{test3apx}
Under assumptions  \emph{({H.5-6})} and the null hypothesis  $\bs{H_0^{m}}$, one has 
$$(Y_n^{m})^2
\xrightarrow{\cL}
\chi^2(1)$$ 
on $(\overline{\cE},\mathbb{P}_{\overline{\cE}})$;
and under the alternative hypothesis $\bs{H_1^{m}}$, almost surely on $\overline{\cE}$ one has
$$\lim_{n\rightarrow\infty}(Y_n^{m})^2 = +\infty.$$
\end{theorem}

\noindent\textit{Proof} Let $g_{m}$ be the function defined from $\dR^8$ onto $\dR$ by 
{$g_{m}(x_1,\ldots,x_8)=2x_1+x_2+x_3+2x_5-x_6-x_7$}
so that $\bs{dg_m}$ is the gradient of $g_m$. Thus, the central limit Theorem~\ref{th:TCL} yields 
$$\sqrt{|\dT_{n-1}^*|}\big(g_m(\wh{\bs{p}}_n)-g_m(\bs{p})\big)
\xrightarrow{\cL}
\cN(0,\bs{dg_m}^t\bs{V}\bs{dg_m})=\cN(0,\bs{\Delta}^m)$$ 
{on }$(\overline{\cE},\mathbb{P}_{\overline{\cE}})$.
Under the null hypothesis $\bs{H_0^{m}}$, $g_m(\bs{p})=0$, so that one has
$$|\dT_{n-1}^*|(\bs{\Delta}^m)^{-1}g_m(\wh{\bs{p}}_n)^2
\xrightarrow{\cL}
\chi^2(1)$$ 
{on }$(\overline{\cE},\mathbb{P}_{\overline{\cE}})$.
Lemma~\ref{lemmaGWapx} and Theorem~\ref{th:CVp} give the almost sure convergence of $\wh{\bs{\Delta}}_n^m$ to $\bs{\Delta}^m$. Hence Slutsky's Lemma yields the expected result.
Under the alternative hypothesis $\bs{H_1^{m}}$, one has 
$$Y_n^m = (\wh{\bs{\Delta}}_n^m)^{-1/2}\big(\sqrt{|\dT_{n-1}^*|}\big(g_m(\wh{\bs{p}}_n)-g_m(\bs{p})\big)+\sqrt{|\dT_{n-1}^*|}g_m(\bs{p})\big).$$
The first term converges to a centered normal law and the second term tends to infinity as $|\dT_{n-1}^*|$ tends to infinity a.s. on $(\overline{\cE},\mathbb{P}_{\overline{\cE}})$.
\hspace{\stretch{1}}$ \Box$\\

Our next test compares the reproduction probability vectors of mother cells of type $0$ and $1$.
\begin{itemize}
\item $\bs{H_0^{p}}$: $\bs{p}^{(0)}=\bs{p}^{(1)}$ the symmetry hypothesis,
\item $\bs{H_1^{p}}$: $\bs{p}^{(0)}\neq \bs{p}^{(1)}$ the alternative hypothesis.
\end{itemize}
Let $(\bs{Y}_n^{p})^t\bs{Y}_n^{p}$ be the test statistic defined by
\begin{equation*}
\bs{Y}_n^{p}=|\dT_{n-1}^*|^{1/2}(\wh{\Delta}_n^{p})^{-1/2}(\wh{\bs{p}}^{(0)}-\wh{\bs{p}}^{(1)}),
\end{equation*}
where
$\wh{\Delta}_n^{p} = |\dT_{n-1}^*|\bs{dg}_{p}^t\wh{\bs{V}}_n\bs{dg}_{p}$ and 
$
\bs{dg}_{p}=\left(\begin{array}{r}
\rI_4\\
-\rI_4
\end{array}\right)$.
This test statistic has the following asymptotic properties.
\begin{theorem}
\label{test4apx}
Under assumptions  \emph{({H.5-6})} and the null hypothesis  $\bs{H_0^{p}}$, one has 
$$(\bs{Y}_n^{p})^t\bs{Y}_n^{p}
\xrightarrow{\cL}
\chi^2(4)$$ 
on $(\overline{\cE},\mathbb{P}_{\overline{\cE}})$; and under the alternative hypothesis $\bs{H_1^{p}}$, almost surely on $\overline{\cE}$ one has
$$\lim_{n\rightarrow\infty}\|\bs{Y}_n^{p}\|^2 = +\infty.$$
\end{theorem}

\noindent\textit{Proof} We mimic the proof of Theorem~\ref{test3apx} with $g_{p}$  the function defined from $\dR^8$ onto $\dR^4$ by 
$g_{p}(x_1,\ldots,x_8)=(x_1-x_5,x_2-x_6,x_3-x_7,x_4-x_8)^t$,
so that $\bs{dg}_{p}$ is the gradient of $g_{p}$. 
\hspace{\stretch{1}}$ \Box$
%
\section{Convergence of estimators for the BAR process}
\label{ss:apxBAR}
We now prove the convergence of the estimators for the BAR process, that is the parts of Theorems~\ref{th:CVp} and \ref{th:TCL} concerning $\wh{\bs{\theta}}_n$,  $\wh{{\sigma}}_{n,{i}}^2$ and $\wh{{\rho}}_n$, together with additional technical results, especially the convergence of higher moment estimators required to estimate the asymptotic variances.
\subsection{Preliminary results: laws of large numbers}
\label{ss:apxBARprelim}
In this section, we want to study the asymptotic behavior of various sums of observed data. Most of the results are directly taken from \cite{SGM11}. All external references in this section refer to that paper that will not be cited each time. However, we need additional results concerning higher moments of the BAR process in order to obtain the consistency 
of $\wh{\tau}^4_{{i},n}$ and $\wh{\nu}_n^2$, as there is no such result in \cite{SGM11}. We also give all the explicit formulas so that the interested reader can actually compute the various asymptotic variances. 

Again, our work relies on the strong law of large numbers for square integrable martingales. To ensure that the increasing processes of our martingales are at most $\cO(\pi^n)$ we first need the following lemma.
\begin{lemma}\label{limX8}
Under assumptions \emph{({H.0-6})}, for all $i\in\{0,1\}$ one has
\begin{equation*}
\sum_{j=1}^m\sum_{k\in\dT_n}\delta_{(j,2k+i)}X_{(j,k)}^8=\cO(\pi^n)\qquad\text{a.s.}
\end{equation*}
\end{lemma}
\noindent\textit{Proof} The proof follows the same lines as that of Lemma~6.1. The constants before the terms $A^i_n$, $B^i_n$ and $C^i_n$ therein are replaced respectively by $(4/(1-\beta))^7$, $\alpha^8(4/(1-\beta))^7$ and $2^8$; in the term $A^i_n$, $\veps^2$ is replaced by $\veps^8$; in the term $C^i_n$, $\beta^{2r_k}$ is replaced by $\beta^{8r_k}$; the term $B^i_n$ is unchanged. In the expression of $\dE[(Y^i_{\ell,p})^2]$, one just needs to replace $\tau^4$ by $\mu^{16}_{{i}}$, $\sigma^4$ by $\gamma^{16}_{{i}}$ and $\nu^2\tau^4$ by $\eta^8$. Note that the various moments of the noise sequence are defined in assumption (H.1). The rest of the proof is unchanged.
 \hspace{\stretch{1}}$ \Box$\\

We also state some laws of large numbers for the noise processes.
\begin{lemma}\label{delta-eps}
Under assumptions \emph{({H.0-6})}, for all $i\in\{0,1\}$ and for all integers $0\leq q\leq 4$, one has
\begin{equation*}
\frac{1}{\pi^n}\sum_{j=1}^m\sum_{k\in\dT_{n-1}}\delta_{(j,2k+i)}\veps^q_{(j,2k+i)}=\frac{\pi}{\pi-1}Wz^{i}\dE[\veps_{(1,2+i)}^q]\qquad\text{a.s.}
\end{equation*}
\end{lemma}
\noindent\textit{Proof} This is also a direct consequence of \cite{SGM11} thanks to Lemmas~\ref{pi-T} and \ref{lemLLNtree}. Lemma~5.3 provides the result for $q=0$, Lemma 5.5 for $q=1$, Corollary 5.6 for $q=2$ and Lemma 5.7 for $q=4$. The result for $q=3$ is obtained similarly.
 \hspace{\stretch{1}}$ \Box$\\

In view of these new stronger results, we can now state our first laws of large numbers for the observed BAR process. For $i \in \{0,1\}$ and all integers $1\leq q\leq 4$  let us now define 
\begin{eqnarray*}
H_n^i (q) & =&  \sum_{j=1}^mH_{j,n}^i(q)=\sum_{j=1}^m\sum_{k\in\dT_n}\delta_{(j,2k+i)}X_{(j,k)}^{q},\\
H_n^{01} (q)  &= & \sum_{j=1}^mH_{j,n}^{01}(q)=\sum_{j=1}^m\sum_{k\in\dT_n}\delta_{(j,2k)}\delta_{(j,2k+1)}X_{(j,k)}^{q},
\end{eqnarray*}
and  $\bs{H}_n(q)=(H_n^0(q),H_n^1(q))^t$. 
\begin{lemma}\label{LGN Xq}
Under assumptions \emph{({H.0-6})} and for all integers $1\leq q\leq 4$, one has the following a.s. limits on the non-extinction set $\overline{\cE}$
\begin{eqnarray*}
\lim_{n\rightarrow \infty} \ind{|\dG_n^*|>0}|\dT^*_n|^{-1}{\bs{H}_{n}(q)}&=&\bs{h}(q)=(\rI_2 - \widetilde{\bs{P}}_q)^{-1}\bs{P}^t\wt{\bs{h}}(q),\\
\lim_{n\rightarrow \infty} \ind{|\dG_n^*|>0}|\dT^*_n|^{-1}{{H}_{n}^{01}(q)}&=&h^{01}(q)\ =\ p^{(0)}(1,1)\Big(\wt{{h}}^0(q)+{b_0^q}\frac{h^0(q)}{\pi}\Big)\\
&&\qquad\qquad+p^{(1)}(1,1)\Big(\wt{{h}}^1(q)+{b_1^q}\frac{h^1(q)}{\pi}\Big),
\end{eqnarray*}
where 
\begin{equation*}
\widetilde{\bs{P}}_q  =  {\pi}^{-1}{\bs{P}^t}\left(\begin{array}{cc}
{b_0^q}&0\\0&{b_1^q}
\end{array}\right),\qquad
\bs{h}(q)=\left(\begin{array}{c}h^0(q)\\h^1(q)\end{array}\right),\qquad
\wt{\bs{h}}(q)=\left(\begin{array}{c}\wt{h}^0(q)\\\wt{h}^1(q)\end{array}\right),
\end{equation*}
and {for $i\in\{0,1\}$}
\begin{eqnarray*}
\wt{h}^{{i}}(1)&=&{a_i}z^{{i}},\\
\wt{h}^{{i}}(2)&=&({a}_{{i}}^2+\sigma^2_{{i}})z^{{i}}+2{a}_{{i}}{b}_{{i}}{h^{{i}}(1)}{\pi}^{-1},\\
\wt{h}^{{i}}(3)&=&({a}_{{i}}^3+3{a}_{{i}}\sigma^2_{{i}}+\lambda_{{i}})z^{{i}}+3{b}_{{i}}({a}_{{i}}^2+\sigma^2_{{i}}){h^{{i}}(1)}{\pi}^{-1}+3{a}_{{i}}{b}_{{i}}^2{h^{{i}}}(2){\pi}^{-1},\\
\wt{h}^{{i}}(4)&=&({a}_{{i}}^4+6{a}_{{i}}^2\sigma^2_{{i}}+4{a}_{{i}}\lambda_{{i}}+\tau^4_{{i}})z^{{i}}+4{b}_{{i}}({a}_{{i}}^3+3{a}_{{i}}\sigma^2_{{i}}+\lambda_{{i}}){h^{{i}}(1)}{\pi}^{-1}\\
&&+6{b}_{{i}}^2({a}_{{i}}^2+\sigma^2_{{i}}){h^{{i}}(2)}{\pi}^{-1}+4{a}_{{i}}{b}_{{i}}^3{h^{{i}}(3)}{\pi}^{-1}.
\end{eqnarray*}
\end{lemma}
\noindent\textit{Proof} The results for $q=1$ and $q=2$ come from Propositions~6.3, 6.5 and 6.6 together with Lemma~\ref{lemLLNtree}. The proofs for $q\geq 3$ follow the same lines, using Lemma~\ref{delta-eps} when required and Lemma~\ref{limX8} to bound the increasing processes of the various martingales at stake.  
\hspace{\stretch{1}}$ \Box$\\

To prove the consistency of our estimators, we also need some additional families of laws of large numbers.
\begin{lemma}\label{LGN Xeps}
Under assumptions \emph{({H.0-6})}, for $i\in\{0,1\}$ and for all integers $1\leq p+q \leq 4$, one has the following a.s. limits 
\begin{equation*}
\ind{|\dG_{n}^*|>0}|\dT^*_n|^{-1}\sum_{j=1}^m\sum_{k\in\dT_n}\delta_{(j,2k+i)}X_{(j,k)}^p\veps_{(j,2k+i)}^{q}=\dE[\veps^q_{2+i}]h^i(p)\indnex.
\end{equation*}
\end{lemma}
\noindent\textit{Proof} The proof is similar to that of Theorem~\ref{th:CVp}. For all $1\leq j\leq m$, one has
\begin{eqnarray*}
\lefteqn{\sum_{k\in\dT_n}\delta_{(j,2k+i)}X_{(j,k)}^p\veps_{(j,2k+i)}^{q}}\\
&=&\sum_{\ell=0}^n\sum_{k\in\dG_{\ell}}\delta_{(j,2k+i)}X_{(j,k)}^p\big(\veps^q_{(j,2k+i)}-\dE[\veps^q_{(j,2k+i)}\ |\ \cF_{j,\ell}^{\cO}]\big)+\dE[\veps^q_{2+i}]\sum_{k\in\dT_n}\delta_{(j,2k+i)}X_{(j,k)}^p,
\end{eqnarray*}
as the conditional moment of $\veps_{2k+i}$ are constants by assumption (H.1). The first term is a square integrable $(\cF_{j,n}^{\cO})$-martingale and its increasing process is $\cO(\pi^n)$ thanks to Lemma~\ref{limX8}, thus the first term is $o(\pi^n)$. The limit of the second term is given by Lemma~\ref{LGN Xq}.
\hspace{\stretch{1}}$ \Box$

\begin{lemma}\label{LGN deltaX}
Under assumptions \emph{({H.0-6})}, for $i\in\{0,1\}$ and for all integers $1\leq q\leq 4$, one has the following a.s. limits 
\begin{equation*}
\ind{|\dG_{n}^*|>0}|\dT^*_n|^{-1}\sum_{j=1}^m\sum_{k\in\dT_n}\delta_{(j,2k+i)}X_{(j,2k+i)}^{q}=\big(\pi\wt{h}^i(q)+{b}_{{i}}^qh^i(q)\big)\indnex.
\end{equation*}
\end{lemma}
\noindent\textit{Proof} The proof is obtained by replacing $X_{(j,2k+i)}$ by ${a_i}+{b_i}X_k+\veps_{2k+i}$. One then develops the exponent and uses Lemmas~\ref{lemmaGWapx}, \ref{delta-eps}, \ref{LGN Xq} and \ref{LGN Xeps} to conclude.
\hspace{\stretch{1}}$ \Box$

\begin{lemma}\label{LGN deltaXX}
Under assumptions \emph{({H.0-6})}, for $i\in\{0,1\}$ and for all integers $1\leq p+q\leq 4$, one has the following a.s. limits 
\begin{equation*}
\ind{|\dG_{n}^*|>0}|\dT^*_n|^{-1}\sum_{j=1}^m\sum_{k\in\dT_n}\delta_{(j,2k+i)}X_{(j,k)}^pX_{(j,2k+i)}^{q}=h^i(p,q)\indnex,
\end{equation*}
with
\begin{eqnarray*}
{h}^{{i}}(p,1)&=&a_{{i}}h^{{i}}(p)+b_{{i}}h^{{i}}(p+1),\\
{h}^{{i}}(p,2)&=&(a^2_{{i}}+\sigma^2_{{i}})h^{{i}}(p)+2a_{{i}}b_{{i}}{h^{{i}}(p+1)}+b^2_{{i}}h^{{i}}(p+2),\\
{h}^{{i}}(p,3)&=&(a_{{i}}^3+3a_{{i}}\sigma^2_{{i}}+\lambda_{{i}})h^{{i}}(p)+3b_{{i}}(a^2_{{i}}+\sigma^2_{{i}}){h^{{i}}(p+1)}+3a_{{i}}b^2_{{i}}{h^{{i}}}(p+2)+b^3_{{i}}h^{{i}}(p+3),\\
\end{eqnarray*}
where we used the convention $h^i(0)=z^i\pi$.
\end{lemma}
\noindent\textit{Proof} As above, the proof is obtained by replacing $X_{(j,2k+i)}$ and developing the exponents. Then one uses Lemmas~\ref{lemmaGWapx}, \ref{delta-eps}, \ref{LGN Xq} and \ref{LGN Xeps} to compute the limits.
\hspace{\stretch{1}}$ \Box$\\

\begin{lemma}\label{delta01}
Under assumptions \emph{({H.5-6})}, one has the following a.s. limit
\begin{equation*}
\ind{|\dG_{n}^*|>0}|\dT^*_{n}|^{-1}\sum_{j=1}^m\sum_{k\in\dT_{n}}\delta_{(j,2(2k+i))}\delta_{(j,2(2k+i)+1)}=p^{(i)}(1,1)z^i\pi\indnex.
\end{equation*}
\end{lemma}
\noindent\textit{Proof} First note that $\delta_{(j,2(2k+i))}\delta_{(j,2(2k+i)+1)}=\delta_{(j,2k+i)}\delta_{(j,2(2k+i))}\delta_{(j,2(2k+i)+1)}$. The proof is then similar to that of Theorem~\ref{th:CVp}. One adds and subtract $p^{(i)}(1,1)$ so that a martingale similar to $(M_n)$ naturally appears. The limit of the remaining term is given by Lemma~\ref{lemmaGWapx}.
\hspace{\stretch{1}}$ \Box$

\begin{lemma}\label{LGN Xeps01}
Under assumptions \emph{({H.0-6})}, for all integers $0\leq p+q+r \leq 4$, one has the following a.s. limits 
\begin{equation*}
\ind{|\dG_{n}^*|>0}|\dT^*_n|^{-1}\sum_{j=1}^m\sum_{k\in\dT_n}\delta_{(j,2k)}\delta_{(j,2k+1)}X_{(j,k)}^p\veps_{(j,2k)}^{q}\veps_{(j,2k+1)}^{r}=\dE[\veps^q_{2}\veps^r_3]h^{01}(p)\indnex,
\end{equation*}
where we used the convention $h^{01}(0)=p^{(0)}(1,1)z^0+p^{(1)}(1,1)z^1$.
\end{lemma}
\noindent\textit{Proof} The proof is similar to Lemma~\ref{LGN Xeps}, one adds and subtracts the constant $\dE[\veps^q_{(j,2k)}\veps_{(j,2k+1)}^{r}\ |\ \cF_{j,\ell}^{\cO}]$.
\hspace{\stretch{1}}$ \Box$

\begin{lemma}\label{LGN deltaXXX}
Under assumptions \emph{({H.0-6})}, for all integers $1\leq p+q+r\leq 4$, one has the following a.s. limits 
\begin{equation*}
\ind{|\dG_{n}^*|>0}|\dT^*_n|^{-1}\sum_{j=1}^m\sum_{k\in\dT_n}\delta_{(j,2k)}\delta_{(j,2k+1)}X_{(j,k)}^pX_{(j,2k)}^{q}X_{(j,2k+1)}^{r}=h^{01}(p,q,r)\indnex,
\end{equation*}
with
\begin{eqnarray*}
{h}^{01}(p,1,0)&=&a_{{0}}h^{01}(p)+b_{{0}}h^{01}(p+1),\qquad
{h}^{01}(p,0,1)\ =\ {a}_{{1}}h^{01}(p)+{b}_{{1}}h^{01}(p+1)),\\
{h}^{01}(p,2,0)&=&(a^2_{{0}}+\sigma^2_{{0}})h^{01}(p)+2a_{{0}}b_{{0}}{h^{01}(p+1)}+b^2_{{0}}h^{01}(p+2),\\
{h}^{01}(p,0,2)& =& ({a}_{{1}}^2+\sigma^2_{{1}})h^{01}(p)+2{a}_{{1}}{b}_{{1}}{h^{01}(p+1)}+{b}_{{1}}^2h^{01}1(p+2),\\
{h}^{01}(p,3,0)&=&(a^3_{{0}}+3a_{{0}}\sigma^2_{{0}}+\lambda_{{0}})h^{01}(p)+3b_{{0}}(a^2_{{0}}+\sigma^2_{{0}}){h^{01}(p+1)}+3a_{{0}}
b^2_{{0}}h^{01}(p+2)+b^3_{{0}}h^{01}(p+3),\\
{h}^{01}(p,0,3)&=&({a}_{{1}}^3+3{a}_{{1}}\sigma^2_{{1}}+\lambda_{{1}})h^{01}(p)+3{b}_{{1}}({a}_{{1}}^2+\sigma^2_{{1}})h^{01}(p+1)+3{a}_{{1}}{b}_{{1}}^2h^{01}(p+2)+{b}_{{1}}^3h^{01}(p+3),\\
h^{01}(p,1,1)&=&(a_{{0}}{a}_{{1}}+\rho)h^{01}(p)+(a_{{0}}{b}_{{1}}+b_{{0}}{a}_{{1}})h^{01}(p+1)+b_{{0}}{b}_{{1}}h^{01}(p+2),\\
h^{01}(p,2,1)&=&((a^2_{{0}}+\sigma^2_{{0}}){a}_{{1}}+2a_{{0}}\rho+\alpha)h^{01}(p)+((a^2_{{0}}+\sigma^2_{{0}}){b}_{{1}}+2(a_{{0}}{a}_{{1}}+\rho)b_{{0}})h^{01}(p+1)\\
&&+b_{{0}}({{ 2}}a_{{0}}{b}_{{1}}+b_{{0}}{a}_{{1}})h^{01}(p+2)+b^2_{{0}}{b}_{{1}}h^{01}(p+3),\\
h^{01}(p,1,2)&=&(({a}_{{1}}^2+\sigma^2_{{1}})a_{{0}}+2{a}_{{1}}\rho+\beta)h^{01}(p)+(({a}_{{1}}^2+\sigma^2_{{1}})b_{{0}}+2({a}_{{0}}{a}_{{1}}+\rho){b}_{{1}})h^{01}(p+1)\\
&&+{b}_{{1}}(a_{{0}}{b}_{{1}}+{{ 2}}b_{{0}}{a}_{{1}})h^{01}(p+2)+b_{{0}}{b}_{{1}}^2h^{01}(p+3),\\
h^{01}(0,2,2)&=&(a^2_{{0}}{a}_{{1}}^2+a^2_{{0}}\sigma^2_{{1}}+{a}_{{1}}^2\sigma^2_{{0}}+\nu^2+2a_{{0}}\beta+2{a}_{{1}}\alpha+4a_{{0}}{a}_{{1}}\rho)h^{01}(0)\\
&&+2(b_{{0}}(a_{{0}}({a}_{{1}}^2+\sigma^2_{{1}})+\beta+2{a}_{{1}}\rho)+{b}_{{1}}({a}_{{1}}(a^2_{{0}}+\sigma^2_{{0}})+\alpha+2a_{{0}}\rho))h^{01}(1)\\
&&(b^2_{{0}}({a}_{{1}}^2+\sigma^2_{{1}})+{b}_{{1}}^2(a^2_{{0}}+\sigma^2_{{0}})+4b_{{0}}{b}_{{1}}(a_{{0}}{a}_{{1}}+\rho))h^{01}(2)\\
&&+2b_{{0}}{b}_{{1}}(a_{{0}}{b}_{{1}}+b_{{0}}{a}_{{1}})h^{01}(3)+b^2_{{0}}{b}_{{1}}^2h^{01}(4).
\end{eqnarray*}
\end{lemma}
\noindent\textit{Proof} The proof is obtained by replacing $X_{(j,2k{+i})}$ by $a_{{i}}+b_{{i}}X_k+\veps_{2k{+i}}$ and developing the exponents. One uses Lemmas~\ref{LGN Xq} and \ref{LGN Xeps01} to compute the limits.
\hspace{\stretch{1}}$ \Box$\\

To conclude this section, we prove the convergence of the normalizing matrices $\bs{S}^0_{n}$, $\bs{S}^1_{n}$ and $\bs{S}^{01}_{n}$ where
$$\bs{S}^{01}_{n} =\sum_{j=1}^m  \sum_{k \in \mathbb{T}_{n}}\delta_{(j,2k)}\delta_{(j,2k+1)}\left(
\begin{array}{cc}
1 & X_{(j,k)} \\
X_{(j,k)} &X^2_{(j,k)}
\end{array}\right),$$
with the sum taken over all observed cells that have observed daughters of both types.

\begin{lemma}
\label{mainlemmaapx}
Suppose that assumptions~\emph{({H.0-6})} are satisfied. Then, there exist definite positive matrices $\bs{L}^0$, $\bs{L}^1$ and $\bs{L}^{01}$ such that for $i\in\{0,1\}$ one has 
$$\lim_{n\rightarrow \infty} \ind{|\dG_n^*|>0}|\dT_n^*|^{-1}\bs{S}^i_{n} = \indnex\bs{L}^i,\quad
\lim_{n\rightarrow \infty} \ind{|\dG_n^*|>0}|\dT_n^*|^{-1}{\bs{S}^{01}_{n}} = \indnex\bs{L}^{01}\quad\text{a.s.}$$
where
\begin{equation*}
\bs{L}^i=\left(\begin{array}{cc}
h^i(0)&h^i(1)\\
h^i(1)&h^i(2)
\end{array}\right),\qquad
\bs{L}^{01}=\left(\begin{array}{cc}
h^{01}(0)&h^{01}(1)\\
h^{01}(1)&h^{01}(2)
\end{array}\right).
\end{equation*}
\end{lemma}

\noindent\textit{Proof}
This is a direct consequence of Lemmas~\ref{lemmaGWapx} and \ref{LGN Xq}. 
\hfill$\Box$
%
\subsection{Strong consistency for the estimators of the BAR process}
\label{ss:apxBARconsist}
We could obtain the convergences of our estimators by sharp martingales results as in \cite{SGM11}, see also~\ref{ss:apxGWconsist}. However, we chose the direct approach here. Indeed, our convergences are now direct consequences of the laws of large numbers given in~\ref{ss:apxBARprelim}.\\

\noindent\textit{Proof of Theorem~\ref{th:CVp}, convergence of $\wh{\bs{\theta}}_n$} This is a direct consequence of Lemmas~\ref{mainlemmaapx} and \ref{LGN deltaXX}. Indeed, by Lemma~\ref{LGN deltaXX} one has
\begin{equation*}
\frac{\ind{|\dG_{n-1}^*|>0}}{|\dT^*_{n-1}|}\bs{\Sigma}_{n-1}\wh{\bs{\theta}}_n
=\frac{\ind{|\dG_{n-1}^*|>0}}{|\dT^*_{n-1}|}\sum_{j=1}^m\sum_{k \in \mathbb{T}_{n-1}}
\left(
\begin{array}{c}
\delta_{(j,2k)}X_{(j,2k)}  \\
\delta_{(j,2k)}X_{(j,k)}X_{(j,2k)} \\
\delta_{(j,2k+1)}X_{(j,2k+1)} \\
\delta_{(j,2k+1)}X_{(j,k)}X_{(j,2k+1)}
\end{array}\right)
\xrightarrow[n\rightarrow\infty]{}\left(
\begin{array}{cc}
\bs{L}^0&0\\
0&\bs{L}^1
\end{array}\right)\bs{\theta}\indnex.
\end{equation*}
And one concludes using Lemma~\ref{mainlemmaapx}.
\hfill$\Box$\\

\noindent\textit{Proof of Theorem~\ref{th:CVp}, convergence of $\wh{{\sigma}}_{ {i},n}^2$ and $\wh{{\rho}}_n$} This result is not as direct as the preceding one because of the presence of the $\wh{\veps}_k$ in the various estimators. Take for instance the estimator $\wh{\sigma}^2_{{i},n}$. For all $1\leq j\leq m$, one has
\begin{eqnarray*}
\sum_{k\in\dT_{n-1}}\!\!\!\!\wh{\veps}_{(j,2k{+i})}^2
&=&\sum_{\ell=0}^{n-1}\sum_{k\in\dG_{\ell}}\delta_{(j,2k{+i})}(X_{(j,2k{+i})}-\wh{a}_{{i},\ell}-\wh{b}_{{i},\ell}X_{(j,k)})^2\\
&=&\sum_{k\in\dT_{n-1}}\delta_{(j,2k{+i})}X_{(j,2k{+i})}^2+\sum_{j=1}^m\sum_{\ell=0}^{n-1}\wh{a}_{{i},\ell}^2\sum_{k\in\dG_{\ell}}\delta_{(j,2k{+i})}\\
&&+2\sum_{\ell=0}^{n-1}\wh{a}_{{i},\ell}\wh{b}_{{i},\ell}\sum_{k\in\dG_{\ell}}\delta_{(j,2k{+i})}X_{(j,k)}+\sum_{\ell=0}^{n-1}\wh{b}_{{i},\ell}^2\sum_{k\in\dG_{\ell}}\delta_{(j,2k{+i})}X_{(j,k)}^2\\
&&-2\sum_{\ell=0}^{n-1}\wh{a}_{{i},\ell}\sum_{k\in\dG_{\ell}}\delta_{(j,2k{+i})}X_{(j,2k{+i})}-2\sum_{\ell=0}^{n-1}\wh{b}_{{i},\ell}\sum_{k\in\dG_{\ell}}\delta_{(j,2k{+i})}X_{(j,k)}X_{(j,2k{+i})}.
\end{eqnarray*}
Let us study the limit of the last term. One has
\begin{equation*}
\frac{1}{\pi^n}\sum_{\ell=0}^{n-1}\wh{b}_{{i},\ell}\sum_{k\in\dG_{\ell}}\delta_{(j,2k{+i})}X_{(j,k)}X_{(j,2k{+i})}
=\frac{1}{\pi}\sum_{\ell=0}^{n-1}\frac{1}{\pi^{n-1-\ell}}\left(\wh{b}_{{i},\ell}\frac{1}{\pi^{\ell}}\sum_{k\in\dG_{\ell}}\delta_{(j,2k{+i})}X_{(j,k)}X_{(j,2k{+i})}\right).
\end{equation*}
We now use Lemma~\ref{lemtoepbar} with $\bs{A}_n=\pi^{-n}$ and $\bs{X}_n=\wh{b}_{{i},n}\pi^{-n}\sum_{k\in\dG_{n}}\delta_{(j,2k{+i})}X_{(j,k)}X_{(j,2k{+i})}$. We know from Lemma~\ref{LGN deltaXX} together with Lemma~\ref{lemTG} that $\pi^{-n}\sum_{k\in\dG_{n}}\delta_{(j,2k{+i})}X_{(j,k)}X_{(j,2k{+i})}$ converges to $h^0(1,1)W_j$, and the previous proof gives the convergence of $\wh{b}_{{i},n}$. Thus, one obtains
\begin{equation*}
\frac{1}{\pi^n}\sum_{\ell=0}^{n-1}\wh{b}_{{i},\ell}\sum_{k\in\dG_{\ell}}\delta_{(j,2k{+i})}X_{(j,k)}X_{(j,2k{+i})}
\xrightarrow[n\rightarrow\infty]{}\frac{\pi^2}{\pi-1}W_jbh^0(1,1).
\end{equation*}
We deal with the other terms in the decomposition of the sum of $\wh{\veps}_{2k}^2$ in a similar way, using either Lemma~\ref{LGN Xq}, \ref{LGN deltaX} or \ref{LGN deltaXX}.
Finally, one obtains the almost sure limit on $\overline{\cE}$
\begin{eqnarray*}
\wh{\sigma}^2_{{i},n}&\xrightarrow[n\rightarrow\infty]{}&\big(\wt{h}^{{i}}(2)+b^2_{{i}}h^{{i}}(2)\pi^{-1}+a^2_{{i}} z^{{i}}+2a_{{i}}b_{{i}}h^{{i}}(1)\pi^{-1}-2a_{{i}}\big(\wt{h}^{{i}}(1)+b^2_{{i}}h^{{i}}(1)\pi^{-1}\big)\big)(z^{{i}})^{-1}\\
&&=\sigma^2_{{i}}.
\end{eqnarray*}
To obtain the convergence of $\wh{\rho}_n$ the approach is similar, using the convergence results given in Lemmas~\ref{LGN Xq}, \ref{delta01}, \ref{LGN Xeps01} and \ref{LGN deltaXXX}.
\hfill$\Box$\\

\begin{theorem}
\label{th:consistency2apx}
Under assumptions \emph{({H.0-6})},
$\wh{\tau}^4_{{i},n}$ and $\wh{\nu}^2_n$ converge almost surely to $\tau^4_{{i}}$   and $\nu^2$ respectively on $\overline{\cE}$.
\end{theorem}

\noindent\textit{Proof}
We work exactly along the same lines as the previous proof with higher powers.
\hfill$\Box$
%
\subsection{Asymptotic normality for the estimators of the BAR process}
\label{ss:apxBARnorm}
We first give the asymptotic normality for $\wh{\bs{\theta}}_n$.\\

\noindent\textit{Proof of Theorem~\ref{th:TCL} for $\wh{\bs{\theta}}_n$} 
Define the $4\times4$ matrices
\begin{equation}\label{defGamma}
\bs{\Sigma} = \left(\begin{array}{cc}\bs{L}^0&0\\0&\bs{L}^1\end{array}\right), \qquad\bs{\Gamma} = \left(\begin{array}{cc}\sigma^2_{{0}}\bs{L}^0&\rho\bs{L}^{01}\\\rho\bs{L}^{01}&\sigma^2_{{1}}\bs{L}^1\end{array}\right),\qquad\bs{\Gamma_{\theta}}=\bs{\Sigma}^{-1}\bs{\Gamma} \bs{\Sigma}^{-1}.
\end{equation}
We now follow the same lines as the proof of the first part of Theorem~\ref{th:TCL} with a different filtration. This time we use the observed sister pair-wise filtration defined as follows. For $0 \leq j \leq m$ and $p\geq 0$, let 
 \begin{equation}
  \label{filtG}
 \mathcal{G}^{\cO}_{j,p}=\cO_j \vee \sigma\{\delta_{(j,1)}X_{(j,1)},\ (\delta_{(j,2k)}X_{(j,2k)}, \delta_{(j,2k+1)}X_{(j,2k+1)}),\ 1\leq k\leq p\}
 \end{equation}
 be the $\sigma$-field generated by the $j$-th GW tree  and all the pairs of observed sister cells in genealogy $j$ up to the daughters of cell $(j,p)$, and let 
$\mathcal{G}^{\cO}_{p}=\vee_{j=1}^m \mathcal{G}^{\cO}_{j,p}$
be the $\sigma$-field generated by the union  of all $\mathcal{G}^{\cO}_{j,p}$ for $1\leq j\leq m$.
 Hence, for instance, $(\delta_{(j,2k)}\veps_{(j,2k)}, \delta_{(j,2k+1)}\veps_{(j,2k+1)})$ is $\mathcal{G}^{\cO}_k$-measurable for all $j$.
In addition, assumptions {({H.1})} and  {({H.4-5})} imply that the process
$$(\delta_{(j,2k)}\veps_{(j,2k)}, X_{(j,k)}\delta_{(j,2k)}\veps_{(j,2k)}, \delta_{(j,2k+1)}\veps_{(j,2k+1)}, X_{(j,k)}\delta_{(j,2k+1)}\veps_{(j,2k+1)})^t$$
is a $(\mathcal{G}^{\cO}_k)$-martingale difference sequence.
Indeed, as the non-extinction set $\overline{\cE}$ is in $\mathcal{G}^{\cO}_{k}$ for every $k \geq 1$, it is first easy to prove that 
${\mathbb{E}_{\overline{\cE}}[\delta_{(j,2k)}\veps_{(j,2k)}|\mathcal{G}^{\cO}_{k-1}]} = \mathbb{E}[\delta_{(j,2k)}\veps_{(j,2k)} |\mathcal{G}^{\cO}_{k-1}]$. Then,
for $k \in \dG_n$, using repeatedly the independence properties, one has
\begin{eqnarray*}
\lefteqn{\mathbb{E}[\delta_{(j,2k)}\veps_{(j,2k)} |\mathcal{G}^{\cO}_{k-1}]}\\
&=& \delta_{(j,2k)}\mathbb{E}\big[\mathbb{E}[\veps_{(j,2k)} |\cO \vee \mathcal{F}_n \vee \sigma(\veps_{j,p}, 1 \le j \le m, p \in \dG_{n+1}, p \le 2k-1)] \ \big|\ \mathcal{G}^{\cO}_{k-1}\big]\\
&=&\delta_{(j,2k)}\mathbb{E}\big[\mathbb{E}[\veps_{(j,2k)} |\mathcal{F}_n \vee \sigma(\veps_{j,p}, 1 \le j \le m, p \in \dG_{n+1}, p \le 2k-1)] \ \big|\ \mathcal{G}^{\cO}_{k-1}\big]\\
&=&\delta_{(j,2k)}\mathbb{E}\big[\mathbb{E}[\veps_{(j,2k)} |\mathcal{F}_n] \ \big|\ \mathcal{G}^{\cO}_{k-1}\big]\ =\ \delta_{(j,2k)}\mathbb{E}\big[\mathbb{E}[\veps_{(j,2k)} |\mathcal{F}_{j,n}] \ \big|\ \mathcal{G}^{\cO}_{k-1}\big]\ =\ 0.
\end{eqnarray*}
We introduce a sequence of  $(\mathcal{G}^{\cO}_k)$-mar\-tin\-ga\-les $(\bs{M}^{(n)}_p)_{\{p \geq 1\}} $ defined for all $n,p\geq 1$ by $\bs{M}^{(n)}_p = {|\dT^*_n|}^{-1/2} \sum_{k=1}^{p}\bs{D}_k$, with
\begin{equation*}
\bs{D}_k =\sum_{j=1}^m\bs{D}_{(j,k)}=\sum_{j=1}^m\left( \begin{array}{cccc}
\delta_{(j,2k)}\veps_{(j,2k)}  \\
X_{(j,k)}\delta_{(j,2k)}\veps_{(j,2k)} \\
\delta_{(j,2k+1)}\veps_{(j,2k+1)} \\
X_{(j,k)}\delta_{(j,2k+1)}\veps_{(j,2k+1)}
\end{array}\right).
\end{equation*}
We also introduce the sequence of stopping times $\nu_n=|\dT_n|=2^{n+1}-1$. We are interested in the convergence of the {process} $\bs{M}^{(n)}_{\nu_n}=  {{|\dT^*_n|}}^{-1/2} \sum_{k=1}^{|\dT_n|}\bs{D}_k$. Again, it is easy to prove that 
\begin{eqnarray*}
{\mathbb{E}_{\overline{\cE}}[\bs{D}_k\bs{D}_k^t|\mathcal{G}^{\cO}_{k-1}]}
&=&\mathbb{E}[\bs{D}_k\bs{D}_k^t|\mathcal{G}^{\cO}_{k-1}]
\ = \ \sum_{j=1}^m\left(\begin{array}{cc}
\sigma^2_{{0}}\bs{\varphi}_{(j,k)}^0&\rho\bs{\varphi}_{(j,k)}^{01} \\
\rho\bs{\varphi}_{(j,k)}^{01}  &  \sigma^2_{{1}}\bs{\varphi}_{(j,k)}^1\end{array}\right),
\end{eqnarray*} 
where for $i\in\{0,1\}$,
\begin{equation*}
\bs{\varphi}_{(j,k)}^i=\delta_{(j,2k+i)}\left(\begin{array}{cc}
1&X_{(j,k)}\\
X_{(j,k)}&X_{(j,k)}^2
\end{array}\right),\quad
\bs{\varphi}_{(j,k)}^{01}=\delta_{(j,2k)}\delta_{(j,2k+1)}\left(\begin{array}{cc}
1&X_{(j,k)}\\
X_{(j,k)}&X_{(j,k)}^2
\end{array}\right).
\end{equation*}
Lemma~\ref{mainlemmaapx} yields that  the $\mathbb{P}_{\overline{\cE}}$ almost sure limit of the process
$<\bs{M}^{(n)}>_{\nu_n}= {|\dT^*_n|}^{-1} \sum_{k \in \dT_n}\mathbb{E}_{\overline{\cE}}[\bs{D}_k\bs{D}^t_k|\mathcal{G}^{\cO}_{k-1}]$ is $\bs{\Gamma}$, as
\begin{equation*}
\sum_{k \in \dT_n}\mathbb{E}_{\overline{\cE}}[\bs{D}_k\bs{D}^t_k|\mathcal{G}^{\cO}_{k-1}] =\left(\begin{array}{cc}
\sigma^2_{{0}}\bs{S}_{n}^0&\rho\bs{S}_{n}^{01}\\
\rho\bs{S}_{n}^{01}&\sigma^2_{{1}}\bs{S}_{n}^1
\end{array}\right).
\end{equation*}
Therefore, the assumption A.1 of Theorem~\ref{TCL:Dufloapx} holds under
$\mathbb{P}_{\overline{\cE}}$.  Thanks to  assumptions {({H.1})} and  {({H.4-5})} we can easily  prove that for some $r>2$, one has $\sup_{k\geq 0}
\dE[\|\bs{D}_k\|^r|\mathcal{G}^{\cO}_{k-1}]<\infty$ {a.s.} which in turn implies the Lindeberg condition A.2. We can now conclude that under
$\mathbb{P}_{\overline{\cE}}$ one has 
$${{|\dT^*_{n-1}|}}^{-1/2}\sum_{k \in \mathbb{T}^*_{n-1}}\bs{D}_k
\xrightarrow{\cL}
\cN(0,\bs{\Gamma}).$$
Finally Eq.~(\ref{defLS}) implies that $\sum_{k \in \mathbb{T}^*_{n-1}}\bs{D}_k=\bs{\Sigma}_{n-1}(\bs{\wh{\theta}}_n-\bs{\theta})$. Therefore, the result is a direct consequence of Lemma~\ref{mainlemmaapx} together with Slutsky's Lemma.
\hspace{\stretch{1}}$ \Box$\\

We now turn to the asymptotic normality of $\wh{\sigma}^2_{{i},n}$ and $\wh{\rho}_n$. The direct application of the central limit theorem for martingales to $\wh{\sigma}^2_{{i},n}$ and $\wh{\rho}_n$ is not obvious because of the $\wh{\veps}_{(j,2k+i)}$. We proceed along the same lines as in the proof of the convergence of $\wh{\bs{\sigma}}_{{i},n}^2$, using the decomposition along the generations. However, this time we need a convergence rate for $\wh{\theta}_n$ in order to apply Lemma~\ref{lemtoepbar}. 
\begin{theorem}\label{thmapthetaapx}
Under assumptions \emph{({H.0-6})}, one has
\begin{equation*}
\ind{|\dG_n^*|>0}\| \widehat{\bs{\theta}}_{n}-\bs{\theta} \|^{2}=
\cO \left(\frac{\log |\dT_{n-1}^*|}{|\dT_{n-1}^*|} \right)\indnex
\hspace{1cm}\text{a.s.}
\end{equation*}
\end{theorem}
\noindent\textit{Proof :} This result is based on the asymptotic behavior of the martingale $( \bs{M}_n)$ defined as follows
\begin{equation*}
\bs{M}_n= \sum_{j=1}^m\sum_{k \in \dT_{n-1}}
\left(\begin{array}{c}
\delta_{(j,2k)}\veps_{2j,k},\\
\delta_{(j,2k)}X_{(j,k)}\veps_{(j,2k)},\\
\delta_{(j,2k+1)}\veps_{(j,2k+1)},\\
\delta_{(j,2k+1)}X_{(j,k)}\veps_{(j,2k+1)}
\end{array}\right).
\end{equation*}
For all $n\geq2$, we readily deduce from the definitions of the BAR process and of our estimator $\wh{\bs{\theta}}_n$ that
\begin{equation*}
\wh{\bs{\theta}}_n-\bs{\theta} =\bs{\Sigma}^{-1}_{n-1}
 \sum_{j=1}^m\sum_{k \in \dT_{n-1}}
\left( \begin{array}{cccc}
\delta_{(j,2k)}\veps_{(j,2k)}  \\
\delta_{(j,2k)}X_{(j,k)}\veps_{(j,2k)} \\
\delta_{(j,2k+1)}\veps_{(j,2k+1)} \\
\delta_{(j,2k+1)}X_{(j,k)}\veps_{(j,2k+1)}
\end{array}\right)=\bs{\Sigma}^{-1}_{n-1}\bs{M}_{n}.
\end{equation*}
The sharp asymptotic behavior of $(\bs{M}_n)$ relies on properties of vector martingales. Thanks to Lemma~\ref{lemLLNtree}, the proof follows exactly the same lines as that of the first part of Theorem~3.2 of \cite{SGM11} and is not repeated here.
\hspace{\stretch{1}}$ \Box$\\

We can now turn to the end of the proof of Theorem~\ref{th:TCL} concerning the asymptotic normality of $\bs{\wh{\sigma}}_{{i},n}^2$ and $\bs{\wh{\rho}}_n$.\\

\noindent\textit{Proof of Theorem~\ref{th:TCL}, asymptotic normality of $\bs{\wh{\sigma}}_{{i},n}^2$} Thanks to Eq. (\ref{defbar}) and (\ref{defepschap}), we decompose $\wh{\sigma}^2_{{i},n} - \sigma^2_{{i}}$ into two parts $U_n^{{i}}$ and $V_n^{{i}}$
\begin{eqnarray*}
 |\dT_{n{-1}}^{*{i}}|(\wh{\sigma}^2_{{i},n} - \sigma^2_{{i}}) & = & \sum_{j=1}^m \sum_{\ell=0}^{n-1}\sum_{k \in
\dG_{\ell-1}} \wh{\veps}_{(j,2k{+i})}^2 -{\veps}_{(j,2k{+i})}^2 
+  \sum_{j=1}^m \sum_{\ell=0}^{n-1}\sum_{k \in
\dG_{\ell-1}}   {\delta}_{(j,2k{+i})}({\veps}_{(j,2k{+i})}^2-\sigma^2_{{i}}) \\
& = & \sum_{j=1}^m \sum_{\ell=0}^{n-1}\sum_{k \in\dG_{\ell-1}} u_{(j,k)}^{{i}}
+  \sum_{j=1}^m \sum_{\ell=0}^{n-1}\sum_{k \in\dG_{\ell-1}} v_{(j,k)}^{{i}} = U_n^{{i}} + V_n^{{i}},
\end{eqnarray*}
with
\begin{eqnarray*}
u_{(j,k)}^{{i}} & = &  \delta_{(j,2k{+i})}\big((a_{{i}}-\wh{a}_{{i},\ell})^2 + (b_{{i}}-\wh{b}_{{i},\ell})^2 X_{(j,k)}^2 + 2 (a_{{i}}-\wh{a}_{{i},\ell})  (b_{{i}}-\wh{b}_{{i},\ell}) X_{(j,k)}  \big)\\
 v_{(j,k)}^{{i}} & = &  \delta_{(j,2k{+i})} \big( 2 \big( (a_{{i}}-\wh{a}_{{i},\ell}) + (b_{{i}}-\wh{b}_{{i},\ell}) X_{(j,k)} \big) \veps_{(j,2k{+i})} + {\veps}_{(j,2k{+i})}^2-\sigma^2_{{i}} \big).
\end{eqnarray*}
We first deal with $U_n^{{i}}$ and study the limit of ${\pi^{-n/2}}U_n^{{i}}$. Let us just detail the first term
\begin{eqnarray*} 
\frac{1}{ \pi^{n/2}} \sum_{j=1}^m \sum_{\ell=0}^{n-1}\sum_{k \in\dG_{\ell-1}}   \delta_{(j,2k{+i})} (a_{{i}}-\wh{a}_{{i},\ell})^2 
 &=&   \sum_{\ell=0}^{n-1} \pi^{(\ell-n)/2} {\frac{\ell}{\pi^{\ell/2}}  \frac{(a_{{i}}-\wh{a}_{{i},\ell})^2 }{\ell \pi^{-\ell}}  \Big(\frac{1}{\pi^{\ell}}\sum_{j=1}^m\sum_{k \in\dG_{\ell-1}}  \delta_{(j,2k{+i})} \Big)}\\
& =&  \sum_{\ell=0}^{n-1} \pi^{(\ell-n)/2}x_{{i},\ell}.
\end{eqnarray*}
On the one hand, Lemmas~\ref{lemmaGWapx}, \ref{pi-T} and \ref{lemTG} imply that $\pi^{-\ell} \sum_{k \in\dG_{\ell-1}}  \delta_{(j,2k{+i})}$ converges a.s. to a finite limit. 
On the other hand, thanks to Theorem \ref{thmapthetaapx}, one has ${(a_{{i}}-\wh{a}_{{i},\ell})^2 }(\ell \pi^{-\ell})^{-1} = \cO(1)$ a.s. As a result, one obtains $\lim_{l \rightarrow \infty} x_{{i},\ell}=0$ a.s. as $\pi>1$ by assumption. Therefore, Lemma~\ref{lemtoepbar} yields 
$$\lim_{n \rightarrow \infty} \frac{1}{ \pi^{n/2}} \sum_{j=1}^m \sum_{\ell=0}^{n-1}\sum_{k \in \dG_{l-1}}   \delta_{j,2k{+i}} (a_{{i}}-\wh{a}_{{i},\ell})^2 = 0 \qquad \text{a.s.}.$$
The other terms in $U_n^{{i}}$ are dealt with similarly, using Lemma~\ref{LGN Xq} instead of Lemma~\ref{lemmaGWapx}. One obtains $\lim_{n \rightarrow\infty} \pi^{-n/2}U_n^{{i}} = 0$ a.s. and as a result Lemma~\ref{pi-T} yields $\lim_{n \rightarrow\infty} |\dT_n^{*}|^{-1/2}U_n^{{i}} = 0$.
Let us now deal with the martingale terms $V_n^{{i}}$. {Set $\bs{V}_n=(V_n^{0},V_n^1)^t$.}
Let us remark that ${|\dT^*_n|}^{-1/2}{\bs{V}_n= \bs{M}^{(n)}_{\nu_n}}$ with ${\bs{M}}^{(n)}= ({\bs{M}}^{(n)}_p)_{\{p \geq 1\}} $ 
 the sequence of $\mathcal{G}^{\cO}_p$-{vector }martingales defined by 
$${\bs{M}}^{(n)}_p = {|\dT^*_n|}^{-1/2} \sum_{k=1}^{p}{(v_{k}^0,v_k^1)^t}= {|\dT^*_n|}^{-1/2} \sum_{k=1}^{p}\sum_{j=1}^m{(v_{(j,k)}^0,v_{(j,k)}^1)^t}$$ 
and $\nu_n=2^n-1$ ($\mathcal{G}^{\cO}_p$  defined by (\ref{filtG})).  We want now to apply Theorem \ref{TCL:Dufloapx} to  ${\bs{M}}^{(n)}$. Using Lemmas~\ref{LGN Xq}-\ref{LGN deltaXXX} together with Lemma~\ref{lemtoepbar} and Theorem \ref{thmapthetaapx} along the same lines as above, we obtain the following limit {conditionally to $\overline{\cE}$
\begin{equation*}
\lim_{n\rightarrow\infty}<\bs{M}>_{\nu_n}=\left(\begin{array}{cc}
(\tau_0^4-\sigma_0^4)z^0&(\nu^2-\sigma_0^2\sigma_1^2)h^{01}(0)\pi^{-1}\\
(\nu^2-\sigma_0^2\sigma_1^2)h^{01}(0)\pi^{-1}&(\tau_1^4-\sigma_1^4)z^1
\end{array}\right)=\bs{\Gamma}_{\bs{V}}.
\end{equation*}}
Therefore, assumption A.1 of Theorem~\ref{TCL:Dufloapx} holds under
$\mathbb{P}_{\overline{\cE}}$.  Thanks to  assumptions {({H.1})} and {({H.4-5})} we can prove  that for some $r>2$, $\sup_{k\geq 0}
\dE_{\overline{\cE}}[\|v_k^{{i}}\|^r|\mathcal{G}^{\cO}_{k-1}]<\infty$ a.s.
 which implies the Lindeberg condition. 
Therefore, we obtain that under $\mathbb{P}_{\overline{\cE}}$
\begin{equation*}
|\dT^*_{n}|^{-1/2}{\bs{V}}_n
\xrightarrow{\cL}
\cN(0, {\bs{\Gamma}_{\bs{V}}}).
\end{equation*}
If one sets
{
\begin{equation}\label{defgammasig}
\bs{\Gamma}_{\sigma}=\left(\begin{array}{cc}
(\tau_0^4-\sigma_0^4)(z^0)^{-1}&(\nu^2-\sigma_0^2\sigma_1^2)h^{01}(0)(\pi z^0z^1)^{-1}\\
(\nu^2-\sigma_0^2\sigma_1^2)h^{01}(0)(\pi z^0z^1)^{-1}&(\tau_1^4-\sigma_1^4)(z^1)^{-1}
\end{array}\right),
\end{equation}}
one obtains the expected result {using Slutsky's lemma}.
 \hspace{\stretch{1}}$ \Box$\\

\noindent\textit{Proof of Theorem~\ref{th:TCL}, Asymptotic normality of $\bs{\wh{\rho}}_n$}.
Along the same lines, we show the central limit theorem for $\wh{\rho}_n$. One has
\begin{eqnarray*}
|\dT_{n-1}^{*01}|(\wh{\rho}_n - \rho) 
& \!\!=\!\! &  \sum_{j=1}^m \sum_{\ell=0}^{n-1}\sum_{k \in
\dG_{\ell-1}} (\wh{\veps}_{(j,2k)}    \wh{\varepsilon}_{(j,2k+1)} -{\veps}_{(j,2k)} {\varepsilon}_{(j,2k+1)})\\
&=& \sum_{j=1}^m \sum_{\ell=0}^{n-1}\sum_{k \in
\dG_{\ell-1}} u'_{(j,k)} +  \sum_{j=1}^m \sum_{\ell=0}^{n-1}\sum_{k \in
\dG_{\ell-1}} v'_{(j,k)} = U'_n + V'_n,
\end{eqnarray*}
with
\begin{eqnarray*}
u'_{(j,k)} & = &  \delta_{(j,2k)} \delta_{(j,2k+1)}\big((a_{{0}}-\wh{a}_{{0},\ell})(a_{{1}}-\wh{a}_{{1},\ell}) + (b_{{0}}-\wh{b}_{{0},\ell})(b_{{1}}-\wh{b}_{{1},\ell}) X_{(j,k)}^2.\\
&&+( (a_{{0}}-\wh{a}_{{0}\ell}) (b_{{1}}-\wh{b}_{{1},\ell}) +  (b_{{0}}-\wh{b}_{{0},\ell})(a_{{1}}-\wh{a}_{{1},\ell})) X_{(j,k)}  \big),\\
 v'_{(j,k)} & = &  \delta_{(j,2k)} \delta_{(j,2k+1)} \big(  ( (a_{{0}}-\wh{a}_{{0},\ell}) + (b_{{0}}-\wh{b}_{{0},\ell}) X_{(j,k)} ) \veps_{(j,2k+1)} \\
 &&+ ((a_{{1}}-\wh{a}_{{1},\ell}) + (b_{{1}}-\wh{b}_{{1},\ell}) X_{(j,k)})\veps_{(j,2k)} + {\veps}_{(j,2k)}{\veps}_{(j,2k+1)}-\rho \big).
\end{eqnarray*}
Thanks to Theorem \ref{thmapthetaapx}, it is easy to check that $ \lim_{n \rightarrow \infty }{|\dT^{*01}_{n-1}|}^{1/2}U'_n = 0$ a.s.
Let us define a new sequence of $\mathcal{G}^{\cO}_p$-martingales $(M^{(n)})$  by 
$$M^{(n)}_p = {|\dT^{*01}_{n-1}|}^{-1/2} \sum_{k=1}^{p}v'_{k}={{|\dT_{n-1}^{*01}|}} ^{-1/2}\sum_{k=1}^{p}\sum_{j=1}^mv'{(j,k)}.$$We
clearly have $M^{(n)}_{\nu_n}={|\dT^{*01}_{n-1}|}^{1/2}V'_n$. We obtain the $\mathbb{P}_{\overline{\cE}}$-{ a.s.} limit 
$$\lim_{n\rightarrow\infty}{|\dT^{*01}_{n-1}|}^{-1}\sum_{k\in\dT_n}\dE_{\overline{\cE}}[v_{k}^2\ |\ \cG_{k-1}^{\cO}]= \nu^2-\rho^2.$$So we have assumption A.1 of Theorem~\ref{TCL:Dufloapx}.
We also derive the Lindeberg condition A.2. Consequently, we obtain that under $\mathbb{P}_{\overline{\cE}}$, one has
$$\sqrt{|\dT^{*01}_{n-1}|}V'_n
\xrightarrow{\cL}
\cN(0,\nu^2-\rho^2).$$
Setting
\begin{equation}
\label{defgammarho}
\gamma_{\rho}= \nu^2-\rho^2,
\end{equation}
completes the proof of Theorem~\ref{th:TCL}.
\hspace{\stretch{1}}$ \Box$
%
\subsection{Interval estimation and tests for the BAR process}
\label{ss:CIBAR}
For all $n\geq 1$, define the $4\times4$ matrices $\wh{\bs{\Gamma}}_n$ and $\wh{\bs{\Omega}}_n$ by
\begin{equation*}
\wh{\bs{\Gamma}}_n=|\dT_{n}^*|^{-1}\left(\begin{array}{cc}
\wh{\sigma}^2_{{0},n}\bs{S}_n^0&\wh{\rho}_n\bs{S}_n^{01}\\
\wh{\rho}_n\bs{S}_n^{01}&\wh{\sigma}^2_{{1},n}\bs{S}_n^1
\end{array}\right),\qquad\text{and}\qquad
\wh{\bs{\Omega}}_n=\bs{\Sigma}_n^{-1}\wh{\bs{\Gamma}}_n\bs{\Sigma}_n^{-1}.
\end{equation*}
Note that the matrix $\wh{\bs{\Gamma}}_n$ is the empirical estimator of matrix $\bs{\Gamma}$ while $\wh{\bs{\Omega}}_n$ is the empirical estimator of the asymptotic variance of $\wh{\bs{\theta}}_n-\bs{\theta}$.
\begin{theorem}
\label{th:ICthetaapx}
Under assumptions \emph{({H.0-6})},  for any $0< \epsilon< 1$, the intervals   
\begin{eqnarray*}
\big[\wh{a}_{{0},n}-q_{1-\epsilon/2}{(\wh{\bs{\Omega}}_{n-1}^{1/2})_{1,1}} ;  \wh{a}_{{0},n}+q_{1-\epsilon/2}{(\wh{\bs{\Omega}}_{n-1}^{1/2})_{1,1}}\big],&&
\big[\wh{b}_{{0},n}-q_{1-\epsilon/2}{(\wh{\bs{\Omega}}_{n-1}^{1/2})_{2,2}} ;  \wh{b}_{{0},n}+q_{1-\epsilon/2}{(\wh{\bs{\Omega}}_{n-1}^{1/2})_{2,2}}\big],\\
\big[\wh{a}_{{1},n}-q_{1-\epsilon/2}{(\wh{\bs{\Omega}}_{n-1}^{1/2})_{3,3}} ;  \wh{a}_{{1},n}+q_{1-\epsilon/2}{(\wh{\bs{\Omega}}_{n-1}^{1/2})_{3,3}}\big],&&
\big[\wh{b}_{{1},n}-q_{1-\epsilon/2}{(\wh{\bs{\Omega}}_{n-1}^{1/2})_{4,4}} ;  \wh{b}_{{1},n}+q_{1-\epsilon/2}{(\wh{\bs{\Omega}}_{n-1}^{1/2})_{4,4}}\big]
\end{eqnarray*}
are asymptotic confidence intervals with level $1 - \epsilon$ of the parameters $a_{{0}}$, $b_{{0}}$, $a_{{1}}$ and $b_{{1}}$ respectively.
\end{theorem}

\noindent\textit{Proof}  This is a straightforward consequence of the central limit Theorem~\ref{th:TCL} together with Slutsky's lemma as 
$
\lim_{n\rightarrow\infty}|\dT_{n-1}^*|\wh{\bs{\Omega}}_{n-1}=\bs{\Sigma}^{-1}\bs{\Gamma} \bs{\Sigma}^{-1}
$ 
$\mathbb{P}_{\overline{\cE}}$ a.s. thanks to Lemma~\ref{mainlemmaapx} and Theorem~\ref{th:CVp}.
\hspace{\stretch{1}}$ \Box$\\

Let 
$$\wh{h}^{01}_n(0)=  \wh{p}_n^{(0)}(1,1) |\dT^*_{n}|^{-1}\sum_{j=1}^m\sum_{k\in\dT_{n-1}}\delta_{(j,2k)} + \wh{p}_n^{(1)}(1,1) |\dT^*_{n}|^{-1}\sum_{j=1}^m\sum_{k\in\dT_{n-1}}\delta_{(j,2k+1)}$$ 
be an empirical estimator of $h^{01}(0)$ and
{
\begin{equation*}
\wh{\bs{\Gamma}}_{\sigma,n}=\left(\begin{array}{cc}
(\wh{\tau}_{0,n}^4-\wh{\sigma}_{0,n}^4)\frac{|\dT_n^*|}{|\dT_{n-1}^{*0}|}
&(\wh{\nu}^2_n-\wh{\sigma}_{0,n}^2\wh{\sigma}_{1,n}^2)h^{01}(0)\wh{\pi}_n^{-1}\frac{|\dT_n^*|^2}{|\dT_{n-1}^{*0}||\dT_{n-1}^{*1}|} \\
(\wh{\nu}^2_n-\wh{\sigma}_{0,n}^2\wh{\sigma}_{1,n}^2)h^{01}(0)\wh{\pi}_n^{-1}\frac{|\dT_n^*|^2}{|\dT_{n-1}^{*0}||\dT_{n-1}^{*1}|}
&(\wh{\tau}_{1,n}^4-\wh{\sigma}_{1,n}^4)\frac{|\dT_n^*|}{|\dT_{n-1}^{*1}|}
\end{array}\right),
\end{equation*}}
be an empirical estimator of the variance term in the central limit theorem regarding $\sigma^2_{{i}}$.
\begin{theorem}
\label{th:ICsigmaapx}
Under assumptions \emph{({H.0-6})},  for any $0< \epsilon< 1$,  the intervals
\begin{eqnarray*}
\Big[\wh{\sigma}^2_{{i},n}-q_{1-\epsilon/2}\Big(\frac{{\wh{\bs{\Gamma}}_{\sigma,n}}}{|\dT^{*}_{n}|}\Big)^{1/2}_{{i},{i}} &;&  \wh{\sigma}^2_{{i},n}+q_{1-\epsilon/2}\Big(\frac{{\wh{\bs{\Gamma}}_{\sigma,n}}}{|\dT^{*}_{n}|}\Big)^{1/2}_{{i},{i}} 
\Big],\\
\Big[\wh{\rho}_n-q_{1-\epsilon/2}\Big(\frac{\wh{\nu}_n^2-\wh{\rho}_n^2}{|\dT^{*01}_{n-1}|}\Big)^{1/2} &;&  \wh{\rho}_n+q_{1-\epsilon/2}\Big(\frac{\wh{\nu}_n^2-\wh{\rho}_n^2}{|\dT^{*01}_{n-1}|}\Big)^{1/2}\Big]
\end{eqnarray*}
are asymptotic confidence intervals with level $1 - \epsilon$ of the parameters $\sigma^2_{{i}}$ and $\rho$ respectively.
\end{theorem}

\noindent\textit{Proof}  This is a again straightforward consequence of the central limit Theorem~\ref{th:TCL} together with Slutsky's lemma as 
\begin{equation*}
\lim_{n\rightarrow\infty}{\wh{\bs{\Gamma}}_{\sigma,n}={\bs{\Gamma}}_{\sigma}},
\qquad
\lim_{n\rightarrow\infty}\wh{\nu}_n^2-\wh{\rho}_n^2={\nu}^2-{\rho}^2,
\end{equation*}
$\mathbb{P}_{\overline{\cE}}$ almost surely thanks to Lemma~\ref{lemmaGWapx} and Theorems~\ref{th:CVp} and \ref{th:consistency2apx}.
\hspace{\stretch{1}}$ \Box$\\

We now propose two different symmetry tests for the BAR process based on the central limit Theorem \ref{th:TCL}. The first one compares the couples ${(a_0,b_0)}$ and ${(a_1,b_1)}$. Set 
\begin{itemize}
\item $\bs{H_0^{c}}$: ${(a_0,b_0)}={(a_1,b_1)}$ the symmetry hypothesis,
\item $\bs{H_1^{c}}$: ${(a_0,b_0)}\neq {(a_1,b_1)}$ the alternative hypothesis.
\end{itemize}
Let $(\bs{Y}_n^c)^t\bs{Y}_n^c$ be the test statistic defined by
\begin{equation*}
\bs{Y}_n^c=|\dT_{n-1}^*|^{1/2}(\wh{\bs{\Delta}}_n^c)^{-1/2}({\wh{a}_{0,n}-\wh{a}_{1,n}, \wh{b}_{0,n}-\wh{b}_{1,n}})^t,
\end{equation*}
where
\begin{equation*}
\wh{\bs{\Delta}}_n^c = |\dT^*_{n-1}|\bs{dg_c}^t\wh{\bs{\Omega}}_{n-1}\bs{dg_c},\quad
\bs{dg_c}=\left(\begin{array}{cccc}1&0&-1&0\\0&1&0&-1\end{array}\right)^t.
\end{equation*}
\begin{theorem}
\label{test1apx}
Under assumptions \emph{({H.0-6})} and the null hypothesis  $\bs{H_0^{c}}$ one has
$$(\bs{Y}_n^c)^t\bs{Y}_n^c
\xrightarrow{\cL}
\chi^2(2)$$ on $(\overline{\cE},\mathbb{P}_{\overline{\cE}})$; and under the alternative hypothesis $\bs{H_1^{c}}$, almost surely on $\overline{\cE}$ one has
$$\lim_{n\rightarrow\infty}\|\bs{Y}_n^c\|^2 = +\infty.$$ 
\end{theorem}

\noindent\textit{Proof}  We mimic again the proof of Theorem~\ref{test3apx} with $g_c$ the function defined from $\dR^4$ onto $\dR^2$ by 
$g_c(x_1,x_2,x_3,x_4)=\big(x_1-x_3, x_2-x_4\big)^t$,
so that $\bs{dg_c}$ is the gradient of $g_c$. 
\hspace{\stretch{1}}$ \Box$\\

Our {next} test compares the fixed points ${a_0/(1-b_0)}$ and ${a_1/(1-b_1)}$, {which are the asymptotic means of $X_{(j,2k)}$ and $X_{(j,2k+1)}$ respectively.}
Set
\begin{itemize}
\item $\bs{H_0^{f}}$: ${a_0/(1-b_0)}={a_1/(1-b_1)}$ the symmetry hypothesis,
\item $\bs{H_1^{f}}$: ${a_0/(1-b_0)}\neq {a_1/(1-b_1)}$ the alternative hypothesis.
\end{itemize}
Let $(Y_n^f)^2$ be the test statistic defined by
\begin{equation*}
Y_n^f=|\dT_{n-1}^*|^{1/2}(\wh{\bs{\Delta}}_n^f)^{-1/2}\big(\wh{a}_{{0},n}/(1-\wh{b}_{{0},n})-\wh{a}_{{1},n}/(1-\wh{b}_{{1},n})\big),
\end{equation*}
where $\wh{\bs{\Delta}}_n^f= |\dT^*_{n-1}|\bs{dg_f}^t\wh{\bs{\Omega}}_{n-1}\bs{dg_f}$, and
$\bs{dg_f}=\big(1/(1-\wh{b}_{{0},n}), \wh{a}_{{0},n}/(1-\wh{b}_{{0},n})^2, -1/(1-\wh{b}_{{1},n}), -\wh{a}_{{1},n}/(1-\wh{b}_{{1},n})^2\big)^t.$
This test statistic has the following asymptotic properties.
\begin{theorem}
\label{test2apx}
Under assumptions \emph{({H.0-6})} and the null hypothesis  $\bs{H_0^{f}}$, one has $$(Y_n^f)^2
\xrightarrow{\cL}
\chi^2(1)$$ on $(\overline{\cE},\mathbb{P}_{\overline{\cE}})$;
and under the alternative hypothesis $\bs{H_1^{f}}$, almost surely on $\overline{\cE}$ one has
$$\lim_{n\rightarrow\infty}(Y_n^f)^2 = +\infty.$$ 
\end{theorem}

\noindent\textit{Proof} We mimic again proof of Theorem~\ref{test3apx} with $g_f$  the function defined from $\dR^4$ onto $\dR$ by 
$g_f(x_1,x_2,x_3,x_4)=\big(x_1/(1-x_2)-x_3/(1-x_4)\big)$,
so that $\bs{dg_f}$ is the gradient of $g_f$. 
\hspace{\stretch{1}}$ \Box$\\

{Finally, our last test compares the even and odd variances $\sigma^2_0$ and $\sigma^2_1$ of the noise sequence. Set
\begin{itemize}
\item $\bs{H_0^{\sigma}}$: $\sigma^2_0=\sigma^2_1$ the symmetry hypothesis,
\item $\bs{H_1^{\sigma}}$: $\sigma^2_0\neq \sigma^2_1$ the alternative hypothesis.
\end{itemize}
Let $(Y_n^\sigma)^2$ be the test statistic defined by
\begin{equation*}
Y_n^\sigma=|\dT_{n-1}^*|^{1/2}(\wh{\bs{\Delta}}_n^\sigma)^{-1/2}\big(\wh{\sigma}^2_{0,n}-\wh{\sigma}^2_{1,n})\big),
\end{equation*}
where $\wh{\bs{\Delta}}_n^\sigma= |\dT^*_{n-1}|\bs{dg_\sigma}^t\wh{\bs{\Gamma}}_{\sigma,n-1}\bs{dg_\sigma}$, and
$\bs{dg_\sigma}=(1,-1)^t.$
This test statistic has the following asymptotic properties.
\begin{theorem}
\label{testsigmaapx}
Under assumptions \emph{({H.0-6})} and the null hypothesis  $\bs{H_0^{\sigma}}$, one has $$(Y_n^\sigma)^2
\xrightarrow{\cL}
\chi^2(1)$$ on $(\overline{\cE},\mathbb{P}_{\overline{\cE}})$;
and under the alternative hypothesis $\bs{H_1^{\sigma}}$, almost surely on $\overline{\cE}$ one has
$$\lim_{n\rightarrow\infty}(Y_n^\sigma)^2 = +\infty.$$ 
\end{theorem}
%
\noindent\textit{Proof} We mimic one last time the proof of Theorem~\ref{test3apx} with $g_\sigma$  the function defined from $\dR^2$ onto $\dR$ by 
$g_\sigma(x_1,x_2)=(x_1-x_2)$,
so that $\bs{dg_\sigma}$ is the gradient of $g_\sigma$. 
\hspace{\stretch{1}}$ \Box$}

\end{document}